# Precision measurement of the cross section for $e^+e^- \rightarrow K^+K^-(\gamma)$ with the initial-state radiation method at BABAR

J. P. Lees, V. Poireau, and V. Tisserand
*Laboratoire d'Annecy-le-Vieux de Physique des Particules (LAPP),*
*Université de Savoie, CNRS/IN2P3, F-74941 Annecy-Le-Vieux, France*

E. Grauges
*Universitat de Barcelona, Facultat de Fisica, Departament ECM, E-08028 Barcelona, Spain*

A. Palano[ab]
*INFN Sezione di Bari[a]; Dipartimento di Fisica, Università di Bari[b], I-70126 Bari, Italy*

G. Eigen and B. Stugu
*University of Bergen, Institute of Physics, N-5007 Bergen, Norway*

D. N. Brown, L. T. Kerth, Yu. G. Kolomensky, M. J. Lee, and G. Lynch
*Lawrence Berkeley National Laboratory and University of California, Berkeley, California 94720, USA*

H. Koch and T. Schroeder
*Ruhr Universität Bochum, Institut für Experimentalphysik 1, D-44780 Bochum, Germany*

C. Hearty, T. S. Mattison, J. A. McKenna, and R. Y. So
*University of British Columbia, Vancouver, British Columbia, Canada V6T 1Z1*

A. Khan
*Brunel University, Uxbridge, Middlesex UB8 3PH, United Kingdom*

V. E. Blinov[ac], A. R. Buzykaev[a], V. P. Druzhinin[ab], V. B. Golubev[ab], E. A. Kravchenko[ab], A. P. Onuchin[ac],
S. I. Serednyakov[ab], Yu. I. Skovpen[ab], E. P. Solodov[ab], K. Yu. Todyshev[ab], and A. N. Yushkov[a]
*Budker Institute of Nuclear Physics SB RAS, Novosibirsk 630090[a],*
*Novosibirsk State University, Novosibirsk 630090[b],*
*Novosibirsk State Technical University, Novosibirsk 630092[c], Russia*

D. Kirkby, A. J. Lankford, and M. Mandelkern
*University of California at Irvine, Irvine, California 92697, USA*

B. Dey, J. W. Gary, O. Long, and G. M. Vitug
*University of California at Riverside, Riverside, California 92521, USA*

C. Campagnari, M. Franco Sevilla, T. M. Hong, D. Kovalskyi, J. D. Richman, and C. A. West
*University of California at Santa Barbara, Santa Barbara, California 93106, USA*

A. M. Eisner, W. S. Lockman, A. J. Martinez, B. A. Schumm, and A. Seiden
*University of California at Santa Cruz, Institute for Particle Physics, Santa Cruz, California 95064, USA*

D. S. Chao, C. H. Cheng, B. Echenard, K. T. Flood, D. G. Hitlin, P. Ongmongkolkul, and F. C. Porter
*California Institute of Technology, Pasadena, California 91125, USA*

R. Andreassen, Z. Huard, B. T. Meadows, B. G. Pushpawela, M. D. Sokoloff, and L. Sun
*University of Cincinnati, Cincinnati, Ohio 45221, USA*

P. C. Bloom, W. T. Ford, A. Gaz, U. Nauenberg, J. G. Smith, and S. R. Wagner
*University of Colorado, Boulder, Colorado 80309, USA*

R. Ayad[*] and W. H. Toki
*Colorado State University, Fort Collins, Colorado 80523, USA*

B. Spaan
*Technische Universität Dortmund, Fakultät Physik, D-44221 Dortmund, Germany*

R. Schwierz
*Technische Universität Dresden, Institut für Kern- und Teilchenphysik, D-01062 Dresden, Germany*

D. Bernard and M. Verderi
*Laboratoire Leprince-Ringuet, Ecole Polytechnique, CNRS/IN2P3, F-91128 Palaiseau, France*

S. Playfer
*University of Edinburgh, Edinburgh EH9 3JZ, United Kingdom*

D. Bettoni[a], C. Bozzi[a], R. Calabrese[ab], G. Cibinetto[ab], E. Fioravanti[ab],
I. Garzia[ab], E. Luppi[ab], L. Piemontese[a], and V. Santoro[a]
*INFN Sezione di Ferrara[a]; Dipartimento di Fisica e Scienze della Terra, Università di Ferrara[b], I-44122 Ferrara, Italy*

R. Baldini-Ferroli, A. Calcaterra, R. de Sangro, G. Finocchiaro,
S. Martellotti, P. Patteri, I. M. Peruzzi,[†] M. Piccolo, M. Rama, and A. Zallo
*INFN Laboratori Nazionali di Frascati, I-00044 Frascati, Italy*

R. Contri[ab], E. Guido[ab], M. Lo Vetere[ab], M. R. Monge[ab], S. Passaggio[a], C. Patrignani[ab], and E. Robutti[a]
*INFN Sezione di Genova[a]; Dipartimento di Fisica, Università di Genova[b], I-16146 Genova, Italy*

B. Bhuyan and V. Prasad
*Indian Institute of Technology Guwahati, Guwahati, Assam, 781 039, India*

M. Morii
*Harvard University, Cambridge, Massachusetts 02138, USA*

A. Adametz and U. Uwer
*Universität Heidelberg, Physikalisches Institut, D-69120 Heidelberg, Germany*

H. M. Lacker
*Humboldt-Universität zu Berlin, Institut für Physik, D-12489 Berlin, Germany*

P. D. Dauncey
*Imperial College London, London, SW7 2AZ, United Kingdom*

U. Mallik
*University of Iowa, Iowa City, Iowa 52242, USA*

C. Chen, J. Cochran, W. T. Meyer, S. Prell, and A. E. Rubin
*Iowa State University, Ames, Iowa 50011-3160, USA*

A. V. Gritsan
*Johns Hopkins University, Baltimore, Maryland 21218, USA*

N. Arnaud, M. Davier, D. Derkach, G. Grosdidier, F. Le Diberder, A. M. Lutz,
B. Malaescu,[‡] P. Roudeau, A. Stocchi, L. L. Wang,[§] and G. Wormser
*Laboratoire de l'Accélérateur Linéaire, IN2P3/CNRS et Université Paris-Sud 11,
Centre Scientifique d'Orsay, F-91898 Orsay Cedex, France*

D. J. Lange and D. M. Wright
*Lawrence Livermore National Laboratory, Livermore, California 94550, USA*

J. P. Coleman, J. R. Fry, E. Gabathuler, D. E. Hutchcroft, D. J. Payne, and C. Touramanis
*University of Liverpool, Liverpool L69 7ZE, United Kingdom*

A. J. Bevan, F. Di Lodovico, and R. Sacco
*Queen Mary, University of London, London, E1 4NS, United Kingdom*

G. Cowan
*University of London, Royal Holloway and Bedford New College, Egham, Surrey TW20 0EX, United Kingdom*

J. Bougher, D. N. Brown, and C. L. Davis
*University of Louisville, Louisville, Kentucky 40292, USA*

A. G. Denig, M. Fritsch, W. Gradl, K. Griessinger, A. Hafner, E. Prencipe, and K. Schubert
*Johannes Gutenberg-Universität Mainz, Institut für Kernphysik, D-55099 Mainz, Germany*

R. J. Barlow[¶] and G. D. Lafferty
*University of Manchester, Manchester M13 9PL, United Kingdom*

E. Behn, R. Cenci, B. Hamilton, A. Jawahery, and D. A. Roberts
*University of Maryland, College Park, Maryland 20742, USA*

R. Cowan, D. Dujmic, and G. Sciolla
*Massachusetts Institute of Technology, Laboratory for Nuclear Science, Cambridge, Massachusetts 02139, USA*

R. Cheaib, P. M. Patel,[**] and S. H. Robertson
*McGill University, Montréal, Québec, Canada H3A 2T8*

P. Biassoni[ab], N. Neri[a], and F. Palombo[ab]
*INFN Sezione di Milano[a]; Dipartimento di Fisica, Università di Milano[b], I-20133 Milano, Italy*

L. Cremaldi, R. Godang,[††] P. Sonnek, and D. J. Summers
*University of Mississippi, University, Mississippi 38677, USA*

X. Nguyen, M. Simard, and P. Taras
*Université de Montréal, Physique des Particules, Montréal, Québec, Canada H3C 3J7*

G. De Nardo[ab], D. Monorchio[ab], G. Onorato[ab], and C. Sciacca[ab]
*INFN Sezione di Napoli[a]; Dipartimento di Scienze Fisiche,
Università di Napoli Federico II[b], I-80126 Napoli, Italy*

M. Martinelli and G. Raven
*NIKHEF, National Institute for Nuclear Physics and High Energy Physics, NL-1009 DB Amsterdam, The Netherlands*

C. P. Jessop and J. M. LoSecco
*University of Notre Dame, Notre Dame, Indiana 46556, USA*

K. Honscheid and R. Kass
*Ohio State University, Columbus, Ohio 43210, USA*

J. Brau, R. Frey, N. B. Sinev, D. Strom, and E. Torrence
*University of Oregon, Eugene, Oregon 97403, USA*

E. Feltresi[ab], M. Margoni[ab], M. Morandin[a], M. Posocco[a], M. Rotondo[a], G. Simi[a], F. Simonetto[ab], and R. Stroili[ab]
*INFN Sezione di Padova[a]; Dipartimento di Fisica, Università di Padova[b], I-35131 Padova, Italy*

S. Akar, E. Ben-Haim, M. Bomben, G. R. Bonneaud, H. Briand,
G. Calderini, J. Chauveau, Ph. Leruste, G. Marchiori, J. Ocariz, and S. Sitt
*Laboratoire de Physique Nucléaire et de Hautes Energies,*

*IN2P3/CNRS, Université Pierre et Marie Curie-Paris6,*
*Université Denis Diderot-Paris7, F-75252 Paris, France*

M. Biasini[ab], E. Manoni[a], S. Pacetti[ab], and A. Rossi[a]
*INFN Sezione di Perugia[a]; Dipartimento di Fisica, Università di Perugia[b], I-06123 Perugia, Italy*

C. Angelini[ab], G. Batignani[ab], S. Bettarini[ab], M. Carpinelli[ab],[‡‡] G. Casarosa[ab], A. Cervelli[ab], F. Forti[ab],
M. A. Giorgi[ab], A. Lusiani[ac], B. Oberhof[ab], E. Paoloni[ab], A. Perez[a], G. Rizzo[ab], and J. J. Walsh[a]
*INFN Sezione di Pisa[a]; Dipartimento di Fisica, Università di Pisa[b]; Scuola Normale Superiore di Pisa[c], I-56127 Pisa, Italy*

D. Lopes Pegna, J. Olsen, and A. J. S. Smith
*Princeton University, Princeton, New Jersey 08544, USA*

R. Faccini[ab], F. Ferrarotto[a], F. Ferroni[ab], M. Gaspero[ab], L. Li Gioi[a], and G. Piredda[a]
*INFN Sezione di Roma[a]; Dipartimento di Fisica,*
*Università di Roma La Sapienza[b], I-00185 Roma, Italy*

C. Bünger, O. Grünberg, T. Hartmann, T. Leddig, C. Voß, and R. Waldi
*Universität Rostock, D-18051 Rostock, Germany*

T. Adye, E. O. Olaiya, and F. F. Wilson
*Rutherford Appleton Laboratory, Chilton, Didcot, Oxon, OX11 0QX, United Kingdom*

S. Emery, G. Hamel de Monchenault, G. Vasseur, and Ch. Yèche
*CEA, Irfu, SPP, Centre de Saclay, F-91191 Gif-sur-Yvette, France*

F. Anulli, D. Aston, D. J. Bard, J. F. Benitez, C. Cartaro, M. R. Convery, J. Dorfan, G. P. Dubois-Felsmann,
W. Dunwoodie, M. Ebert, R. C. Field, B. G. Fulsom, A. M. Gabareen, M. T. Graham, C. Hast,
W. R. Innes, P. Kim, M. L. Kocian, D. W. G. S. Leith, P. Lewis, D. Lindemann, B. Lindquist, S. Luitz,
V. Luth, H. L. Lynch, D. B. MacFarlane, D. R. Muller, H. Neal, S. Nelson, M. Perl, T. Pulliam,
B. N. Ratcliff, A. Roodman, A. A. Salnikov, R. H. Schindler, A. Snyder, D. Su, M. K. Sullivan, J. Va'vra,
A. P. Wagner, W. F. Wang, W. J. Wisniewski, M. Wittgen, D. H. Wright, H. W. Wulsin, and V. Ziegler
*SLAC National Accelerator Laboratory, Stanford, California 94309 USA*

W. Park, M. V. Purohit, R. M. White,[§§] and J. R. Wilson
*University of South Carolina, Columbia, South Carolina 29208, USA*

A. Randle-Conde and S. J. Sekula
*Southern Methodist University, Dallas, Texas 75275, USA*

M. Bellis, P. R. Burchat, T. S. Miyashita, and E. M. T. Puccio
*Stanford University, Stanford, California 94305-4060, USA*

M. S. Alam and J. A. Ernst
*State University of New York, Albany, New York 12222, USA*

R. Gorodeisky, N. Guttman, D. R. Peimer, and A. Soffer
*Tel Aviv University, School of Physics and Astronomy, Tel Aviv, 69978, Israel*

S. M. Spanier
*University of Tennessee, Knoxville, Tennessee 37996, USA*

J. L. Ritchie, A. M. Ruland, R. F. Schwitters, and B. C. Wray
*University of Texas at Austin, Austin, Texas 78712, USA*

J. M. Izen and X. C. Lou
*University of Texas at Dallas, Richardson, Texas 75083, USA*

F. Bianchi[ab], F. De Mori[ab], A. Filippi[a], D. Gamba[ab], and S. Zambito[ab]

*INFN Sezione di Torino[a]; Dipartimento di Fisica Sperimentale, Università di Torino[b], I-10125 Torino, Italy*

L. Lanceri[ab] and L. Vitale[ab]

*INFN Sezione di Trieste[a]; Dipartimento di Fisica, Università di Trieste[b], I-34127 Trieste, Italy*

F. Martinez-Vidal, A. Oyanguren, and P. Villanueva-Perez

*IFIC, Universitat de Valencia-CSIC, E-46071 Valencia, Spain*

H. Ahmed, J. Albert, Sw. Banerjee, F. U. Bernlochner, H. H. F. Choi, G. J. King, R. Kowalewski,
M. J. Lewczuk, T. Lueck, I. M. Nugent, J. M. Roney, R. J. Sobie, and N. Tasneem

*University of Victoria, Victoria, British Columbia, Canada V8W 3P6*

T. J. Gershon, P. F. Harrison, and T. E. Latham

*Department of Physics, University of Warwick, Coventry CV4 7AL, United Kingdom*

H. R. Band, S. Dasu, Y. Pan, R. Prepost, and S. L. Wu

*University of Wisconsin, Madison, Wisconsin 53706, USA*

(Dated: June 12, 2013)

A precise measurement of the cross section for the process $e^+e^- \to K^+K^-(\gamma)$ from threshold to an energy of $5\,\mathrm{GeV}$ is obtained with the initial-state radiation (ISR) method using $232\,\mathrm{fb}^{-1}$ of data collected with the *BABAR* detector at $e^+e^-$ center-of-mass energies near $10.6\,\mathrm{GeV}$. The measurement uses the effective ISR luminosity determined from the $e^+e^- \to \mu^+\mu^-(\gamma)\gamma_{\mathrm{ISR}}$ process with the same data set. The corresponding lowest-order contribution to the hadronic vacuum polarization term in the muon magnetic anomaly is found to be $a_\mu^{KK,\mathrm{LO}} = (22.93 \pm 0.18_{\mathrm{stat}} \pm 0.22_{\mathrm{syst}}) \times 10^{-10}$. The charged kaon form factor is extracted and compared to previous results. Its magnitude at large energy significantly exceeds the asymptotic QCD prediction, while the measured slope is consistent with the prediction.

## I. INTRODUCTION

The measurement of the $e^+e^- \to K^+K^-(\gamma)$ cross section presented in this paper takes place in the context of a precision measurement of $R = \sigma(e^+e^- \to \mathrm{hadrons})/\sigma(e^+e^- \to \mu^+\mu^-)$ at low energy. Integrals involving $R$ enter the calculations of the hadronic contribution to vacuum polarization (VP). Uncertainties on VP are a limiting factor in precise comparisons of data with the Standard Model (SM) expectations, such as the value of the muon magnetic moment anomaly $a_\mu$. The analysis makes use of several data-driven techniques to measure efficiencies and constrain systematic uncertainties below the 1% level. Accurate parameters for the $\phi$ resonance are determined and the charged kaon form factor is extracted for the first time in a large energy range, from the $K^+K^-$ production threshold to 5 GeV.

Unlike previous measurements, which were performed through energy scans, the present analysis uses the initial-state radiation (ISR) method [1–4]. The $e^+e^- \to K^+K^-(\gamma)$ cross section at the reduced energy $\sqrt{s'}$ is deduced from the measured spectrum of $e^+e^- \to K^+K^-(\gamma)\gamma_{\mathrm{ISR}}$ events produced at the center-of-mass (c.m.) energy $\sqrt{s}$. The reduced energy is related to the energy $E_\gamma^*$ of the ISR photon in the $e^+e^-$ c.m. frame by $s' = s(1 - 2E_\gamma^*/\sqrt{s})$, and it is equal to the mass $m_{KK}$ of the hadronic final state, or $m_{KK\gamma}$ if an additional photon from final-state radiation (FSR) has been emitted. The cross section for the process $e^+e^- \to K^+K^-(\gamma)$ is related to the $\sqrt{s'}$ spectrum of $e^+e^- \to K^+K^-(\gamma)\gamma_{\mathrm{ISR}}$ events through

$$\frac{dN_{K^+K^-(\gamma)\gamma_{\mathrm{ISR}}}}{d\sqrt{s'}} = \frac{dL_{\mathrm{ISR}}^{\mathrm{eff}}}{d\sqrt{s'}} \, \varepsilon_{KK\gamma}(\sqrt{s'}) \, \sigma_{KK(\gamma)}^0(\sqrt{s'}), \quad (1)$$

where $dL_{\mathrm{ISR}}^{\mathrm{eff}}/d\sqrt{s'}$ is the effective ISR luminosity, $\varepsilon_{KK\gamma}$ is the full acceptance for the event sample, and $\sigma_{KK(\gamma)}^0$ is the 'bare' cross section for the process $e^+e^- \to K^+K^-(\gamma)$ (including final-state radiative effects), from which the

*Now at the University of Tabuk, Tabuk 71491, Saudi Arabia

†Also with Università di Perugia, Dipartimento di Fisica, Perugia, Italy

‡Now at Laboratoire de Physique Nucléaire et de Hautes Energies, IN2P3/CNRS, Paris, France

§Also with Institute of High Energy Physics, Beijing, China

¶Now at the University of Huddersfield, Huddersfield HD1 3DH, UK

**Deceased

††Now at University of South Alabama, Mobile, Alabama 36688, USA

‡‡Also with Università di Sassari, Sassari, Italy

§§Now at Universidad Técnica Federico Santa Maria, Valparaiso, Chile 2390123

leptonic and hadronic vacuum polarization contributions are excluded. In contrast to most measurements based on the ISR method, the effective ISR luminosity does not rely on the theoretical radiator function [1–4], which describes the probability to emit an ISR photon of energy $E_\gamma^*$ in a given angular acceptance, or on the external measurement of the data luminosity. Instead, the effective ISR luminosity is determined from the measurement of the $e^+e^- \to \mu^+\mu^-(\gamma)\gamma_{\mathrm{ISR}}$ spectrum with the same data sample, through a relation similar to Eq. (1) where the $e^+e^- \to \mu^+\mu^-$ cross section is given by Quantum Electrodynamics (QED). In this manner several systematic uncertainties cancel. In particular, the cross section measurement is mostly insensitive to higher-order ISR corrections and other theoretical uncertainties that affect the kaon and muon channels equally. The method used in this analysis has been developed for the precision measurement of the $e^+e^- \to \pi^+\pi^-(\gamma)$ cross section and is expounded in Ref. [5].

This paper is organized as follows. In Sec. II, we describe the data samples used in the analysis and the event selection. In Sec. III, selection efficiencies and the corresponding corrections based on differences between data and Monte Carlo (MC) simulation are presented. Sec. IV describes backgrounds. Sec. V is dedicated to the unfolding of the mass spectrum, while Sec. VI describes the acceptance corrections applied to the cross section. Finally, Sec. VII reports the results for the cross section and kaon form-factor from threshold to 5 GeV, and includes the $K^+K^-$ contribution to the anomalous magnetic moment of the muon.

## II. SAMPLES AND EVENT SELECTION

Signal events are characterized by two charged-particle tracks and a high energy photon, all required to lie within the detector acceptance. In addition, in order to control the overall efficiency to high precision, it is found necessary to include higher-order radiation. The next-to-leading-order (NLO) is sufficient to reach accuracies of $10^{-3}$, so the analysis considers $KK\gamma\gamma$ as well as $KK\gamma$ final states, where the additional photon can be either ISR or FSR.

The data were produced at the SLAC National Accelerator Laboratory at the PEP-II $e^+e^-$ collider, operated at and 40 MeV below the peak of the $\Upsilon(4S)$ resonance, $\sqrt{s} = 10.58$ GeV. The analysis is based on 232 fb$^{-1}$ of data collected with the *BABAR* detector, described in detail in Ref. [6]. Charged-particle tracks are measured with a five-layer double-sided silicon vertex tracker (SVT) together with a 40-layer drift chamber (DCH), both inside a 1.5 T superconducting solenoid magnet. Photons are assumed to originate from the primary vertex defined by the charged tracks of the event, and their energy and position are measured in a CsI(Tl) electromagnetic calorimeter (EMC). Charged-particle identification (PID) uses the ionization energy loss ($\mathrm{d}E/\mathrm{d}x$) in the SVT and DCH, the Cherenkov radiation detected in a ring-imaging device (DIRC), the shower energy deposit in the EMC ($E_{\mathrm{cal}}$), and the shower shape in the instrumented flux return (IFR) of the magnet. The IFR system is composed of modules of resistive-plate chambers interspaced with iron slabs, arranged in a layout with a barrel and two endcaps.

Signal and background ISR processes are simulated with the AfkQed event generator based on Ref. [7]. The signal $KK(\gamma)\gamma_{\mathrm{ISR}}$ sample corresponds to about 30 times the integrated luminosity of the data. The main ISR photon, $\gamma_{\mathrm{ISR}}$, is generated within the angular range $[\theta_{\min}^* = 20°,\ \theta_{\max}^* = 160°]$ in the c.m. system[1], wider than the geometrical acceptance of the detector. Additional radiation from the initial state is generated with the structure function method [8] in the collinear approximation, while additional final-state photons are generated with the PHOTOS [9] program. A minimum-mass requirement $m_{K^+K^-\gamma_{\mathrm{ISR}}} > 8$ GeV/$c^2$, applied at generation, limits the emission of a second hard photon in simulation. Background processes $e^+e^- \to q\overline{q}$ ($q = u,d,s,c$) are generated with the JETSET [10] generator, and $e^+e^- \to \tau^+\tau^-$ with the KORALB [11] program. The response of the *BABAR* detector is simulated using the GEANT4 [12] package. In addition, since the additional ISR generated by AfkQed is inadequate, large samples of Monte Carlo (MC) events at the four-momentum level, dedicated to specific ISR studies, are produced with the nearly-exact NLO Phokhara [13] generator.

### A. Topological selection

Two-charged-particle ISR events are selected by requiring a photon with an energy $E_\gamma^* > 3$ GeV in the $e^+e^-$ c.m. and laboratory polar angle with respect to the $e^-$ beam in the range $[0.35–2.4]$ rad, and exactly two tracks of opposite charge, each with momentum $p > 1$ GeV/$c$ and within the angular range $[0.40–2.45]$ rad. If more than one photon is detected, the ISR photon is assumed to be the candidate with the highest $E_\gamma^*$. The charged-particle tracks are required to have at least 15 hits in the DCH, to originate within 5 mm of the collision axis (distance of closest approach doca$_{xy} < 5$ mm) and within 6 cm from the beam spot along the beam direction ($|\Delta_z| < 6$ cm), and to extrapolate to the DIRC and IFR active areas, in order to exclude low-efficiency regions. Events can be accompanied by any number of reconstructed tracks not satisfying the above criteria, and any number of additional photons. To ensure a rough momentum balance at the preselection level (hereafter called 'preselection cut'), the ISR photon is required to lie within 0.3 rad of the missing momentum of all the tracks (or of the tracks plus the other photons).

---

[1] Unless otherwise stated, starred quantities are measured in the $e^+e^-$ c.m. and un-starred quantities in the laboratory.

## B. Kaon identification

To select $KK\gamma$ candidates, the two tracks are required to be identified as kaons. Kaon identification ($K$-ID) proceeds from an optimization between efficiency and misidentification of particles of other types ($e, \mu, \pi, p$) as kaons. Electron contamination is strongly reduced by a criterion based on a combination of $E_{\rm cal}$ and $dE/dx$. In addition, kaons are positively selected through a likelihood estimator $\mathcal{L}$ based on the $dE/dx$ in the DCH and SVT and on the Cherenkov angle in the DIRC. Tracks whose number of associated photons in the DIRC is not sufficient to define a Cherenkov ring ($N_{\rm DIRC} < 3$) are rejected. Pions and protons are rejected through selection criteria on likelihood ratios: $\mathcal{L}_K/(\mathcal{L}_K + \mathcal{L}_\pi) > 0.9$ and $\mathcal{L}_K/(\mathcal{L}_K + \mathcal{L}_p) > 0.2$, respectively. Kaons are further required to fail muon identification. To maximize the $K$-ID efficiency, the veto against the muon background relies on a tight muon selector, where muons are identified by an energy deposit in the EMC consistent with a minimum ionizing particle (MIP), and topological requirements in the IFR (penetration, number of hits, and shower width). A $K$-ID efficiency of 80% is achieved. The probabilities to misidentify a muon or pion as a kaon are below 10% and are measured in the data, as described in Sec. III C 2. The proton misidentification probability is 5% or less and is taken from simulation.

## C. ISR Kinematic fit with an additional photon

Following the method described in Ref. [5] for the analysis of the $\mu\mu\gamma$ and $\pi\pi\gamma$ processes, the event definition is enlarged to include the radiation of one photon in addition to the already required ISR photon. Two kinematic fits to the $e^+e^- \to KK(\gamma)\gamma_{\rm ISR}$ hypothesis are performed:

- If an additional photon is detected in the EMC with energy $E_\gamma > 20$ MeV, its energy and angles are used in a three-constraint (3C) fit. We call this an 'FSR' fit, although the extra photon can be either from FSR or from ISR at large angle. In case multiple extra photons are detected, the FSR fit is performed using each photon in turn and the fit with the smallest $\chi^2_{\rm FSR}$ is retained.

- For every event, an additional photon from ISR at small angle is assumed to be emitted along either the $e^+$ or the $e^-$ beam direction. The corresponding so-called 2C ISR fit ignores additional photons measured in the EMC and returns the energy $E^*_{\gamma\ {\rm add.ISR}}$ of the fitted collinear ISR photon.

In both cases, the constrained fit uses the measured $\gamma_{\rm ISR}$ direction, and momenta and angles of the two tracks,

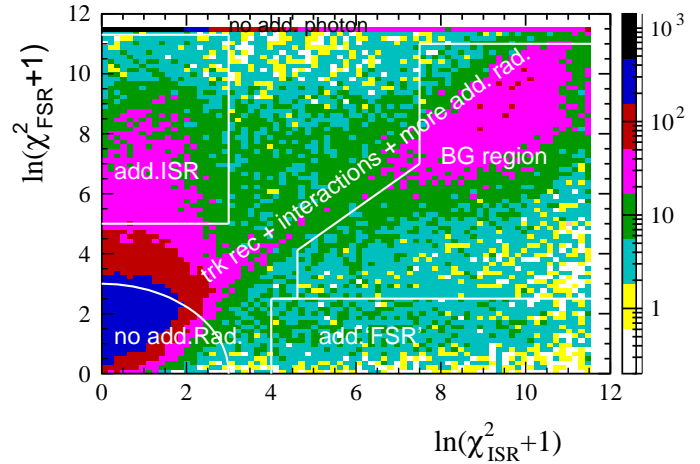

FIG. 1: (color online). The 2D-$\chi^2$ distribution for the $KK(\gamma)\gamma_{\rm ISR}$ data sample in the [0.98–5] GeV/$c^2$ range of the fitted $KK$ mass, where different interesting regions are defined. The line labeled 'no add. photon' corresponds to events with no detected additional photon, which are characterized by the $\chi^2_{\rm ISR}$ value only.

along with their covariance matrix, to solve the four-momentum conservation equations. The kaon mass is assumed for the two charged particles. The energy of the primary ISR photon is not used in either fit. Each event is characterized by the $\chi^2$ values of the two kinematic fits, except for the 12.5% of the candidates with no extra measured photons, for which only the $\chi^2$ from the ISR fit ($\chi^2_{\rm ISR}$) is available. The $K^+K^-$ invariant mass $m_{KK}$ is obtained using the fitted parameters of the two kaons from the ISR fit if $\chi^2_{\rm ISR} < \chi^2_{\rm FSR}$, and from the FSR fit in the reverse case.

Most events appear at small values of both $\chi^2_{\rm ISR}$ and $\chi^2_{\rm FSR}$, as shown on the 2D-$\chi^2$ distribution (Fig. 1), but the tails along the axes clearly indicate events with additional radiation: small-angle ISR along the $\chi^2_{\rm FSR}$ axis (with large fitted photon energies at large values of $\chi^2_{\rm FSR}$), and FSR or large-angle ISR along the $\chi^2_{\rm ISR}$ axis (with large measured photon energies at large values of $\chi^2_{\rm ISR}$). Events along the diagonal do not satisfy either hypothesis and result from either the finite resolution of the kaon track measurement or the direction of the primary ISR photon, or possibly from additional radiation of more than one photon. Events affected by secondary interactions also lie along the diagonal. Multibody background is expected to populate the region where both $\chi^2$ are large, and, consequently, a background ('BG') region is defined in the 2D-$\chi^2$ plane, as indicated in Fig. 1.

For the cross section measurement, the $KK(\gamma)\gamma_{\rm ISR}$ candidates are required to satisfy a 'tight' selection $\ln(\chi^2_{\rm ISR} + 1) < 3$. In order to study efficiencies, backgrounds and mass resolution, we define a 'loose' selection, given by the full 2D-$\chi^2$ plane except for the BG-labeled region. We refer to the region within the loose selection but excluded by the tight selection as the 'intermediate' region.

### D. Raw mass spectrum and angular distribution in the $K^+K^-$ frame

Figure 2 shows the $K^+K^-$ mass spectrum measured in the data with the tight $\chi^2$ selection, without background subtraction or correction for acceptance. The spectrum exhibits distinct features. Besides the prominent $\phi$ resonance at $1.02\,\mathrm{GeV}/c^2$, other structures are visible in the $[1.6-2.5]\,\mathrm{GeV}/c^2$ mass region, as well as signals at the $J/\psi$ and $\psi(2S)$ resonances. These features are examined in Sec. VII.

Since the background is small in the $\phi$ region, as discussed in Sec. IV, one can readily verify that the angular distribution in the $KK$ center-of-mass frame behaves as expected for a decaying vector-particle with helicity one. Figure 3 shows the distributions of the cosine of the angle $\theta_{\gamma\mathrm{trk}}$ between the ISR photon and the charged tracks in the $KK$ center-of-mass frame, for data and MC. The two distributions are consistent with each other and follow the expected $\sin^2\theta_{\gamma\mathrm{trk}}$ shape.

## III. EFFICIENCY AND DATA-MC CORRECTIONS FOR DETECTOR SIMULATION

The mass-dependent overall acceptance $\varepsilon_{KK\gamma}$ is determined with the full AfkQed plus GEANT4 simulation, with corrections applied to account for observed differences between data and MC. Through specific studies, we determine the ratios of the efficiencies $\varepsilon_i$ obtained with the same methods in data and simulation for the trigger, tracking, PID, and $\chi^2$ selection, and we apply them as mass-dependent corrections to the measured $m_{KK}$ spectrum. Corrections to the geometrical acceptance are treated separately in Sec. VI, as most corrections cancel in the $KK(\gamma)$ cross section measurement using the effective luminosity from $\mu\mu(\gamma)\gamma_{\mathrm{ISR}}$ data.

The event efficiency corrected for detector effects is thus

$$\varepsilon = \varepsilon_{\mathrm{MC}}\ \left(\frac{\varepsilon_{\mathrm{trig}}^{\mathrm{data}}}{\varepsilon_{\mathrm{trig}}^{\mathrm{MC}}}\right)\ \left(\frac{\varepsilon_{\mathrm{track}}^{\mathrm{data}}}{\varepsilon_{\mathrm{track}}^{\mathrm{MC}}}\right)\ \left(\frac{\varepsilon_{\mathrm{PID}}^{\mathrm{data}}}{\varepsilon_{\mathrm{PID}}^{\mathrm{MC}}}\right)\ \left(\frac{\varepsilon_{\chi^2}^{\mathrm{data}}}{\varepsilon_{\chi^2}^{\mathrm{MC}}}\right). \quad (2)$$

The mass-dependent corrections $C_i = \left(\varepsilon_i^{\mathrm{data}}/\varepsilon_i^{\mathrm{MC}}\right)$ are discussed below. Most trigger, tracking, and PID inefficiencies arise from a geometrical effect, namely the overlap of the two tracks in the DCH, EMC, or IFR. To avoid correlations between the $C_i$ terms, the efficiencies are determined sequentially, with minimal requirements on the subsequent step. Trigger efficiency is measured on enlarged signal samples selected without a requirement on the actual number of reconstructed tracks. Tracking efficiency is measured with events that have passed the triggers. PID efficiencies and misidentification probabilities are measured with two-track events. Biases associated with the efficiency determination, which result from the measurement method, are studied with MC and are normalized to data through data-to-MC comparison of characteristic distributions once the physics origin of the bias is identified. Since the data sample in the $\phi$ peak region is so pure, efficiencies are measured in the restricted mass range $1.0 < m_{KK} < 1.05\,\mathrm{GeV}/c^2$ and extrapolated to higher mass regions, where large backgrounds preclude direct measurements. Extrapolation is performed using the $KK(\gamma)\gamma_{\mathrm{ISR}}$ MC to sample the corrections $C_i$ determined in the restricted phase space as functions of the relevant variables. Details of the procedure applied to determine each $C_i$ correction term are given below.

### A. Trigger and filter efficiency corrections

Trigger and filter efficiencies are determined in data and MC using complementary triggers. Several sets of criteria (triggers) are applied to each of three levels, hardware (L1), software (L3) and event filter (EF), and the response of each is recorded with the event. In addition, a prescaled sample is retained regardless of whether any trigger is satisfied. The efficiencies of all triggers can therefore be cross calibrated with the others. These are all multipurpose triggers common to BABAR, with none specifically designed to retain two-track ISR events.

Events for the trigger studies are selected through the 1C fit designed for tracking studies (see below) applied to the one- or two-prong sample. The 'primary' track is required to be identified as a kaon with momentum $p > 1\,\mathrm{GeV}/c$, but otherwise minimal requirements are imposed on track quality to avoid correlations with the tracking efficiency measurement.

Inefficiencies of the hardware (L1) and software (L3) triggers are found to be below $10^{-4}$ and $(3.5\pm0.2)\%$, respectively, for data in the vicinity of the $\phi$ peak. They are well reproduced by simulation, and the deviation from unity of the data/MC ratio for the L3 trigger efficiency is found to be $(-0.3\pm1.6)\times10^{-3}$, with no significant variation with $m_{KK}$. The online event filter introduces an inefficiency of $(1.2\pm0.1)\times10^{-3}$ in data, slightly underestimated by MC; a correction of $(0.6\pm0.2)\times10^{-3}$ is applied. Biases on L3 and filter efficiency measurements are observed in MC at a few per mil level. They are due to pairs of non-interacting, minimum-ionizing kaons, whose tracks overlap both in the DCH and EMC. For such events, the tracking-based triggers are degraded while, simultaneously, the triggers based on EMC deposits are enhanced. The biases are calibrated with data using the fractions of double-MIP deposits in the EMC. They are maximal at the $\phi$ mass due to the kinematics of the $\phi$ resonance. The related uncertainties on $C_{\mathrm{trig}}$ are $0.7\times10^{-3}$ under the $\phi$ peak and are extrapolated to about $0.5\times10^{-3}$ at larger masses. At threshold, the uncertainties related to the muon background subtraction in the data sample dominate, and the systematic error on $C_{\text{trig}}$ reaches $1.0 \times 10^{-3}$.

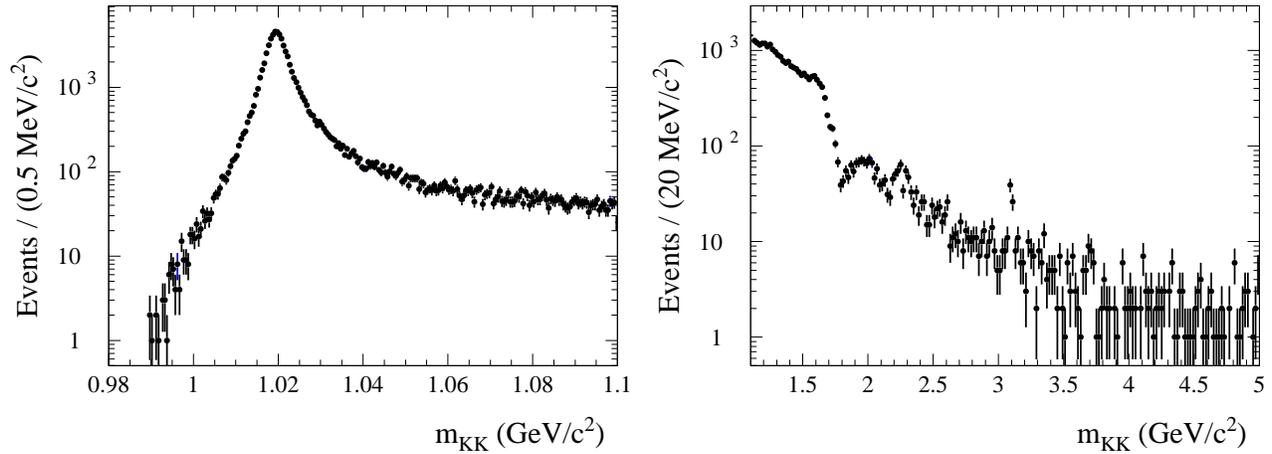

FIG. 2: The $K^+K^-$ invariant mass spectrum for the data sample, after the tight $\chi^2$ selection: $\phi$ mass region (left), masses above $m_\phi$ (right).

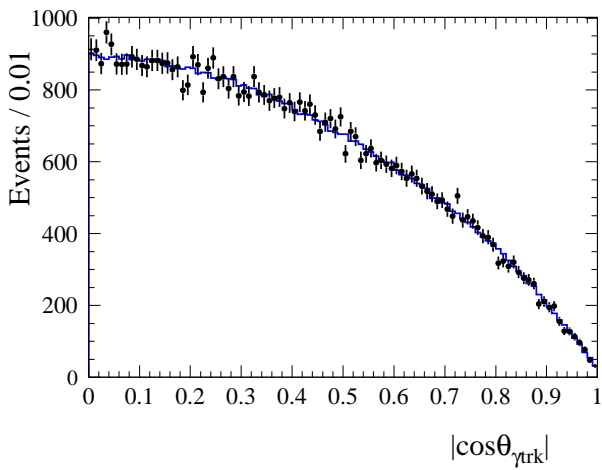

FIG. 3: Distributions of the absolute value of the cosine of the angle between the ISR photon and the charged tracks in the $KK$ center-of-mass, for data (black points) and MC (blue histogram). The $KK$ mass range is from 1.01 to 1.03 GeV/$c^2$. The MC is normalized to the number of events in the data.

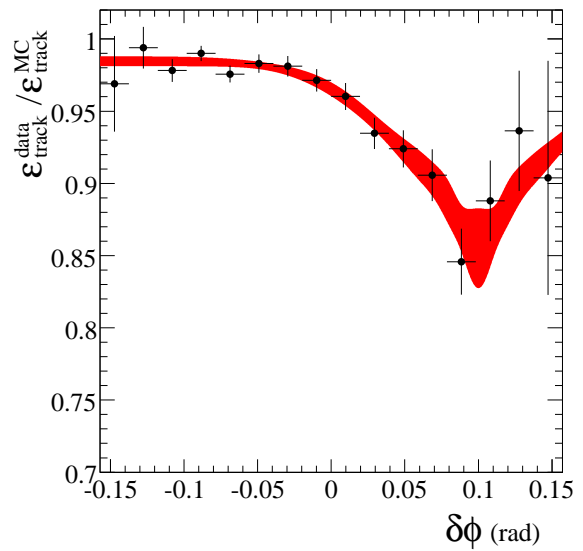

FIG. 4: Fit of the data/MC correction for the tracking efficiency (per event, i.e., for the two tracks) as a function of $\delta\phi$. The function for the fit is a constant plus two Gaussians. The central values of the Gaussians are fixed at 0.1 rad. The red band indicates the errors computed from the covariance matrix of the fit parameters.

## B. Tracking efficiency correction

A 1C kinematic fit is used to select $K^+K^-\gamma_{\text{ISR}}$ events for tracking efficiency studies. The fit is performed on an enlarged tracking sample that includes events with one or two tracks. The fit uses as input only one kaon-identified good track (called 'primary') and the ISR photon, and the momentum vector of the second kaon is predicted from four-momentum conservation. The predicted kaon is required to lie within the tracking acceptance. Only kinematically reconstructed $K^+K^-$ masses in the $\phi$ resonance region ($1.00 < m_{KK} < 1.05$ GeV/$c^2$) are selected in order to reduce the non-kaon background in the tracking sample to the 1% level.

The rate of in-acceptance predicted tracks that are actually reconstructed in the tracking system, with a charge opposite to that of the primary kaon, determines the kaon tracking efficiency. The method yields the intrinsic tracking inefficiency, which is mostly due to interactions in the detector material or kaon decays in flight. In addition to the uncorrelated track loss, a local reduction of the individual track efficiency is induced by the overlap of the tracks in the DCH. The tracking efficiency as a function of the signed angular difference between the positive and negative tracks in the transverse plane $\delta\phi = \phi^+ - \phi^-$ exhibits a dip at small positive values of $\delta\phi$ both in data and MC, which is characteristic of track overlap. This effect has been studied in detail in data and simulation for the $\mu^+\mu^-$ and $\pi^+\pi^-$ final states [5]. The same features are observed for $K^+K^-$, although the $\phi$ mass selection applied to the kaon tracking sample precludes $\delta\phi$ from reaching values larger than $0.15$ rad.

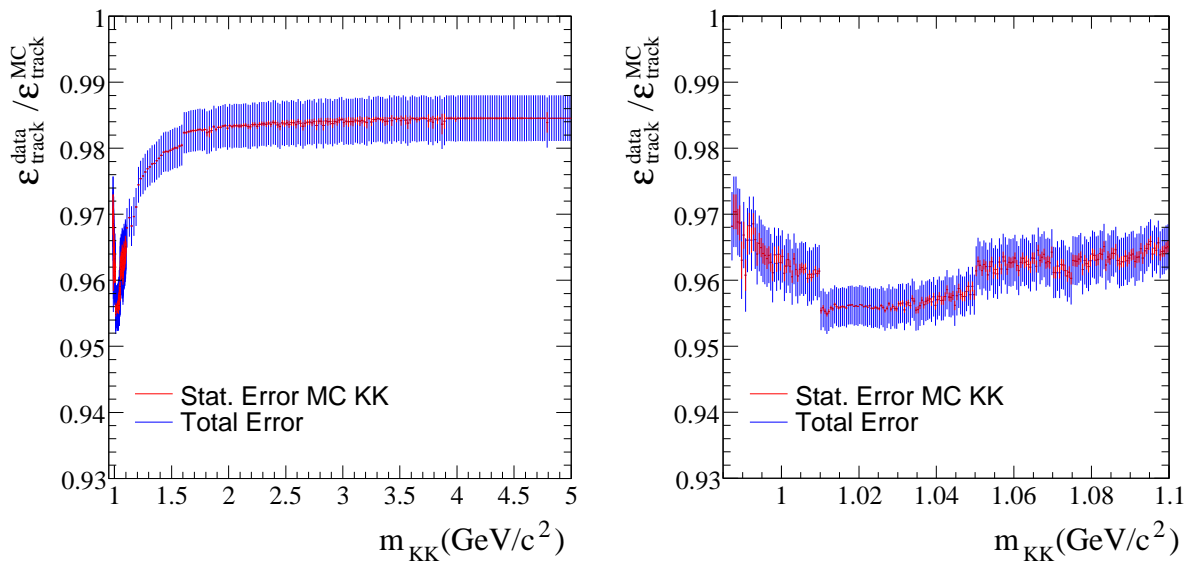

FIG. 5: (color online). The data/MC correction for the tracking-efficiency as a function of $m_{KK}$. The red error bars show the (small) statistical errors from the sampling, whereas the blue ones show the total errors (including the errors from the fit). The figure on the right is a zoom of the figure on the left in the $\phi$ resonance region.

Some difference between data and MC is observed in the magnitude of the effect, as seen in Fig. 4. The $\delta\phi$ dependence of the data/MC correction is fitted with the functional form observed over the full $\delta\phi$ range for muons and pions: besides a flat component due to the intrinsic inefficiency, a double Gaussian is used to describe the sharp asymmetric structure related to the track overlap, located at $\delta\phi \approx 0.1$ rad. As the magnitude of the overlap effect varies with mass, studies of the peak inefficiency are performed with MC on the kaon sample, and, in parallel, on the muon (pion) samples of $\mu\mu\gamma$ ($\pi\pi\gamma$) data and MC events. The general mass dependence of the peak inefficiency is similar for all two-track ISR channels: a maximum of about $1 - 2\%$ around the region of maximum overlap, and a slow decrease to a plateau at higher masses. In the muon sample, where efficiencies can be measured both in data and MC over the full mass range, the data/MC ratio of peak inefficiencies is found to be independent of mass. This validates the extrapolation of the track overlap effect in $KK\gamma$, measured at the $\phi$ mass, to higher masses according to the mass dependence of the peak inefficiency in MC. The latter is obtained in wide mass ranges, and the resulting $C_{track}$ correction is shown in Fig. 5 as a function of mass, where discontinuities reflect the statistical fluctuations of the peak inefficiency values, and errors are fully correlated within the wide mass bins. The correction increases from $3.0\%$ at threshold to about $4.5\%$ in the $\phi$ region, and it decreases to around $1.5\%$ at high masses.

The probability of losing the two tracks in a correlated way, also induced by the track overlap, and the probability for having an extra reconstructed track, are found to be well reproduced by MC in this analysis and small data/MC differences of $0.8 \times 10^{-3}$ and $1.2 \times 10^{-3}$, respectively, are included in the systematic uncertainties. Uncertainties on the bias from the primary-track tagging induce a systematic error of $1.1 \times 10^{-3}$. Together with the uncertainties on the mass dependence of the overlap correction, the dominant contribution to the systematic error is related to the model used to describe the correction as a function of $\delta\phi$. The total systematic uncertainty for the $C_{track}$ correction is smaller than $0.3\%$ below $1.05$ $GeV/c^2$, increasing to about $1\%$ at high mass.

## C. Particle ID efficiency corrections

Separation of ISR two-body processes $e^+e^- \rightarrow x^+x^-(\gamma)\gamma_{ISR}$ ($x = e, \mu, \pi, K, p$) from each other relies on PID. The specific studies conducted to determine the kaon-ID efficiency for data and MC, as well as the $\mu \rightarrow 'K'$ and $\pi \rightarrow 'K'$ misidentification probabilities, are described below. Electron misidentification as a kaon is negligible, as well as data/MC corrections for proton misidentification.

### 1. K-ID efficiency

The method to determine the kaon-ID efficiencies makes use of the two-body ISR sample itself, where one of the produced charged particles is tagged as a kaon and the identification of the second track is probed ('tag-and-probe' method). The PID sample is selected through 1C kinematic fits to the $e^+e^- \to x^+x^-\gamma_{\mathrm{ISR}}$ hypotheses ($x = \mu, \pi, K$) that use only the two charged tracks, with an assigned mass $m_x$, as input. A requirement $\chi^2_{KK} < 15$ is applied to strongly reduce the multihadronic background, as well as a restriction $\chi^2_{KK} < \chi^2_{\pi\pi}$ to reduce the pion contamination. The purity of the kaon-ID sample is further enhanced by requiring the fitted $m_{KK}$ mass to lie in the $\phi$ resonance region. The purity achieved is $(99.0 \pm 0.1)\%$, determined from a fit of the $m_{KK}$ distribution in data, with $\phi$ signal and background shapes taken from MC.

As the efficiency of the muon veto included in the kaon selection varied with time due to degradation in the IFR performance[2], the efficiencies are measured for different data taking periods separately, and combined subsequently. Efficiencies are determined separately for $K^+$ and $K^-$ since a few percent level are observed in the data/MC corrections for $K$-ID efficiency.

The corrections are obtained as a function of the momentum of the charged particle. The restricted mass range of the kaon-ID sample restricts the momentum range of the probed track. The data/MC correction is measured in the [1–5] GeV/$c$ momentum interval and extrapolated to higher momenta through an empirical fit. Sampling of the data/MC corrections obtained for $K^+$ and $K^-$ is performed with the $KK\gamma$ simulation, and results in the $C_{\mathrm{PID}}$ correction shown in Fig. 6. A systematic uncertainty of 0.10% of the correction is assigned to account for the purity of the kaon candidate sample. A systematic uncertainty of 10% is included for each track with a momentum larger than 5 GeV/$c$. The latter uncertainty is negligible for events in the $\phi$ resonance region and becomes important only for events with masses larger than 1.05 GeV/$c^2$.

In addition to the uncorrelated $K$-ID inefficiency measured with the tag-and-probe method, a correlated loss of $K$-ID for both tracks occurs at a rate $f$ due to their overlap, mainly in the DIRC. The $f$ factor is maximum at the $\phi$ mass, where it amounts to $0.0129 \pm 0.0001$ in MC, and vanishes beyond 1.5 GeV/$c^2$. $f$ is measured in data at the $\phi$ resonance with a sample selected irrespective of kaon identification, by fitting the $m_{KK}$ distributions of events with zero, one, or two identified kaons, for the respective number of $\phi$-candidates. The mass-dependence of $f$ is taken from MC. The deviation from unity of the data/MC ratio $(1 - f_{\mathrm{data}})/(1 - f_{\mathrm{MC}})$ amounts to $7 \times 10^{-3}$ at maximum overlap, and vanishes beyond 1.5 GeV/$c^2$. Half the deviation is conservatively added to the $K$-ID systematic uncertainty.

The bias of this method is evaluated with MC, where the number of selected events with two identified kaons, corrected for efficiencies and correlated loss, is compared to the number of events without PID applied. This consistency check includes the extrapolation of the efficiency for track momenta beyond the $\phi$ phase space. Although the mass dependence of the bias indicates that it is also related to overlap effects, no bias larger than $10^{-3}$ is observed at the $\phi$ mass. The full bias is conservatively added to the systematic error.

### 2. $\mu \to$ 'K' and $\pi \to$ 'K' misidentification

The $\mu \to$ 'K' and $\pi \to$ 'K' mis-ID probabilities are determined for MC and data by applying a tag-and-probe method analogous to that used for the $K$-ID efficiency measurement. Pure $\mu\mu\gamma$ and $\pi\pi\gamma$ samples are selected in the restricted mass ranges $m_{\mu\mu} \in [2.5\text{--}5]$ GeV/$c^2$ and $m_{\pi\pi} \in [0.6\text{--}0.9]$ GeV/$c^2$, respectively, to ensure very low contamination of the reference samples from the other two-body ISR channels. Non-two-body ISR event backgrounds are reduced to negligible levels by a tight $\chi^2$ selection on a kinematic fit to the $\mu\mu\gamma$ ($\pi\pi\gamma$) hypothesis. The mass ranges chosen for the $\mu$ and $\pi$ reference samples correspond to regions of maximal contamination to the $KK\gamma$ channel and cover similar angular regions of the detector.

The mis-ID probabilities are determined as a function of the probed-track momentum $p$, and the data/MC mis-ID corrections for $\mu\mu \to$ 'KK' and $\pi\pi \to$ 'KK' are fitted to an empirical function of $p$. The corrections obtained by sampling the above fitted corrections with MC, are shown in Fig. 7. A systematic uncertainty of about 30% of the $\mu\mu \to$ 'KK' correction is estimated by varying the mass range of the reference sample. Even though the data/MC correction is large, it applies to an absolute $\mu\mu \to$ 'KK' mis-ID rate less than 2 per mil and hence induces a negligible systematic uncertainty. For $\pi\pi \to$ 'KK', the mis-ID rate is larger (up to 1%, depending on $m_{KK}$), but the correction is much smaller, and no systematic error is included.

### D. $\chi^2$ efficiency corrections

The measurement of the $\chi^2$ selection efficiency proceeds as in the $\pi\pi(\gamma)$ cross section measurement and we refer to Ref. [5] for the full description. The strategy is to rely on the efficiency measured in the $\mu\mu(\gamma)\gamma_{\mathrm{ISR}}$ anal-ysis to address event losses common to the muon and kaon channels, while the issues specific to kaons are further investigated. Common losses arise because of mis-reconstruction of the ISR photon or tracks and due to additional ISR or higher-order ISR processes. Losses due to additional FSR are restricted to muons as FSR is expected to be very small for kaons, $f_{\text{FSR}}^{KK} \approx f_{\text{FSR}}^{\mu\mu} \cdot \left(\frac{m_\mu}{m_K}\right)^2 = (0.51 \pm 0.02) \times 10^{-3}$, and they are found to be simulated with adequate accuracy (see below). Specific to kaons are interactions in the detector material and decays in flight. The latter are found to be well simulated, and the number of decayed kaons entering the sample is small due to the PID requirements. Event loss due to decays in flight is included in the discussion of interactions below. Other potential differences between the $\chi^2$ selection efficiencies in the muon and kaon channels, such as residual track misreconstruction effects induced by track overlap, are included in the systematic errors.

---

[2] This problem was remedied through IFR detector upgrades, for data collected subsequent to the sample employed for the present analysis [14].

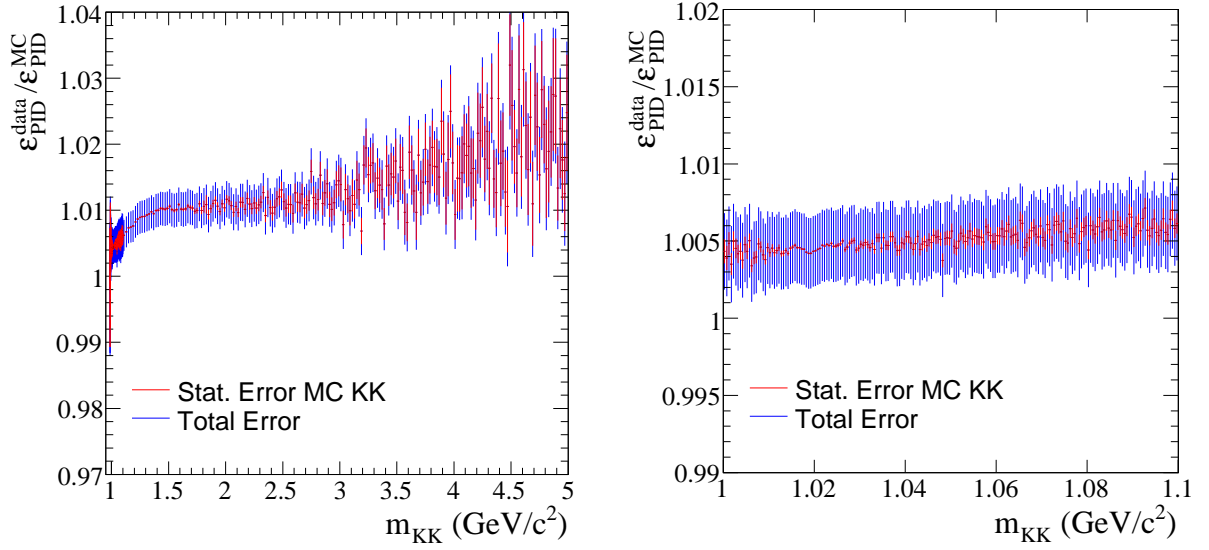

FIG. 6: (color online). The data/MC correction for $K$-ID efficiency as a function of $m_{KK}$. The red error bars show the statistical errors from the sampling, whereas the blue ones show the total errors (including the errors from the fit). The plots correspond to a sampling with MC events in the tight $\chi^2$ region. The figure on the right is a zoom in the $\phi$ resonance region.

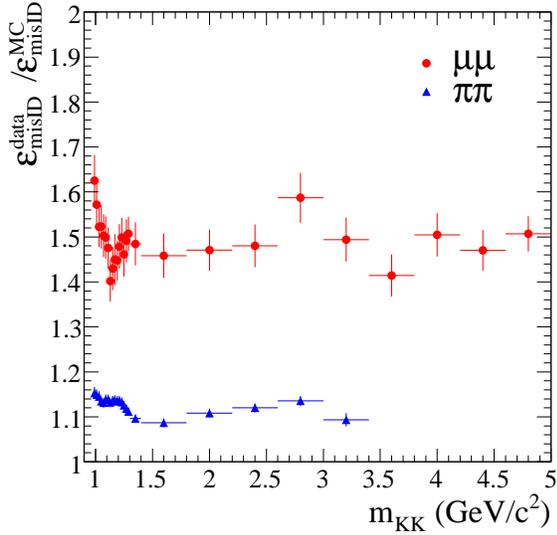

FIG. 7: Data/MC correction for the $\mu\mu \to$ '$KK$' (red points) and $\pi\pi \to$ '$KK$' (blue triangles) mis-ID fractions as a function of $m_{KK}$.

Following the above prescription, the data/MC correction $C_{\chi^2}$ for the $\chi^2$ selection efficiency in the kaon channel is derived from the following expression:

$$C_{\chi^2}(m_{KK}) = C_{\chi^2}^{\mu\mu,\text{FSR sub.}}(m_{KK}) \cdot C_{\chi^2}^{KK,\text{sec.int.}}(m_{KK}), \quad (3)$$

where the first term on the right accounts for the data/MC correction for muons, with the FSR contribution removed, while the correction for kaon secondary interactions is provided by the second term. The data/MC correction for muons, expressed as a function of $m_{KK}$ in Eq. (3), is evaluated at the $m_{\mu\mu}$ mass corresponding to the same track momenta as $m_{KK}$, with assigned mass $m_\mu$ and $m_K$, respectively. In so doing, the $\chi^2$ efficiency is computed for similar kinematical configurations between kaons and muons, especially close to threshold.

### 1. Effects of additional radiation

To assess the validity of the method, we compare the $\chi^2$ distributions in data and MC of events with sizeable additional radiation, either ISR or FSR. The selected kaon samples are restricted to the $\phi$ mass region, and the BG region of the 2D-$\chi^2$ plane is excluded.

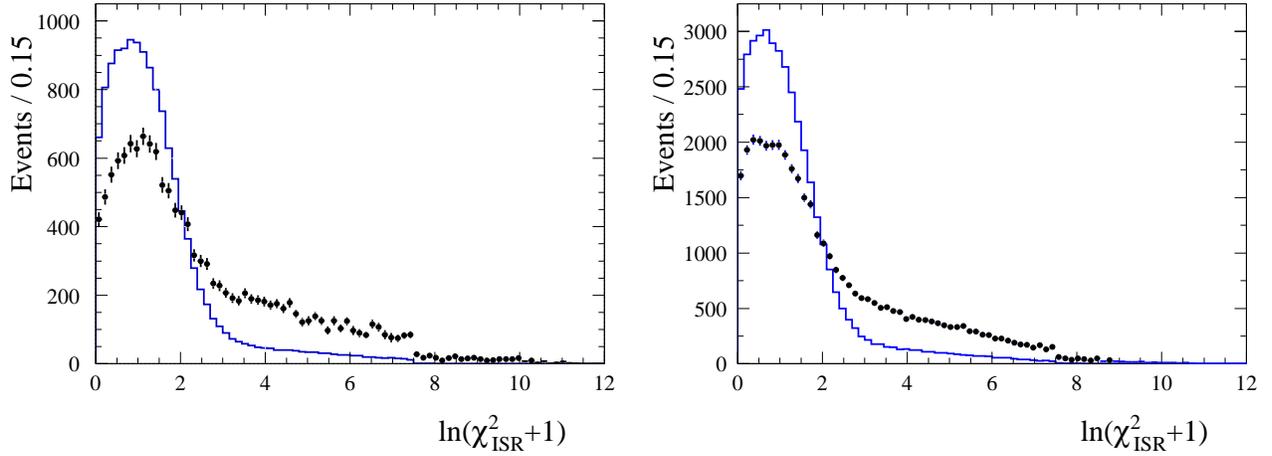

FIG. 8: $\chi^2_{ISR}$ distribution for the kaon ISR subsample (left) and muon ISR subsample (right) in data (points) and MC (histogram). The plots correspond to $0.95 < m_{KK} < 1.1\,\mathrm{GeV}/c^2$ and $m_{\mu\mu} < 1\,\mathrm{GeV}/c^2$, respectively, for events satisfying a loose $\chi^2$ criterion. The MC is normalized to the number of events in the data.

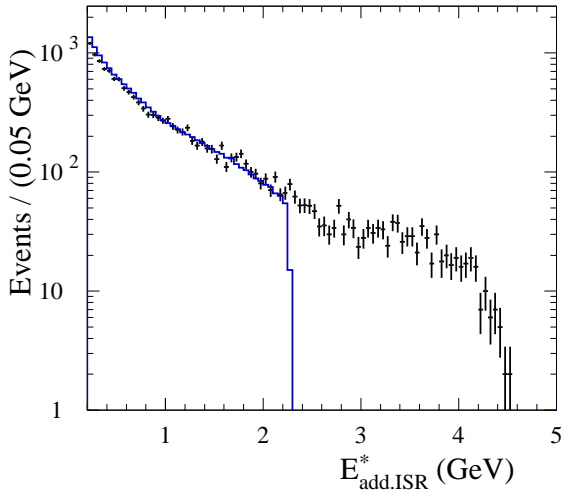

FIG. 9: Energy distribution of the additional ISR photon in the c.m. frame, in the $KK$ ISR subsample in background-subtracted data (points) and MC (histogram). The plot corresponds to $0.95 < m_{KK} < 1.1\,\mathrm{GeV}/c^2$. The MC is normalized to the data luminosity. The sharp cutoff at $2.3\,\mathrm{GeV}$ in MC is caused by the $m_{KK\gamma_{ISR}} > 8\,\mathrm{GeV}/c^2$ requirement set at generation.

For the study of additional ISR at small angles to the beam, we select an 'ISR' subsample by requiring $\ln(\chi^2_{FSR}+1) > \ln(\chi^2_{ISR}+1)$ and $E^*_{\gamma\,add.ISR} > 0.2\,\mathrm{GeV}$. Figure 8 shows the corresponding $\chi^2_{ISR}$ distribution. The data-MC agreement is poor for both muons and kaons because additional ISR is generated by AfkQed in the collinear approximation, while emission at finite angles gives rise to a large high-$\chi^2_{ISR}$ tail in data. Figure 9 shows the c.m. energy distribution of the additional ISR photon, in the ISR subsample. Agreement between data and simulation is observed up to a sharp cutoff at $2.3\,\mathrm{GeV}$ in MC caused by the $m_{KK\gamma_{ISR}} > 8\,\mathrm{GeV}/c^2$ requirement set at generation. However, such a feature is also present in the muon channel and results in a small systematic error on the $\chi^2$ efficiency correction.

For the study of additional FSR and large-angle ISR, we select an 'FSR' subsample by requiring $\ln(\chi^2_{FSR}+1) < \ln(\chi^2_{ISR}+1)$ and $E_{\gamma\,add.FSR} > 0.2\,\mathrm{GeV}$. These events populate the FSR intermediate $\chi^2$ region defined in Sec. II C. The distribution of the angle in the laboratory frame between the additional photon and the closest kaon is shown in Fig. 10. The selected 'FSR' subsample in data is dominated by a large-angle additional ISR signal, which is not present in the AfkQed simulation.

The data (Fig. 10) provide some evidence for FSR photons at angles less than $20°$ with respect to the nearest kaon, as predicted by the MC. The fitted ratio of the rates observed in data and MC is $1.44 \pm 0.95$, for an absolute FSR rate in MC of $5 \times 10^{-4}$. The PHOTOS prescription used in MC to generate FSR is found to describe the data accurately enough. The possible bias of $(0.18 \pm 0.38) \times 10^{-3}$ on the efficiency of the tight $\chi^2$ selection due to incorrect FSR simulation is negligible. PHOTOS is also found to accurately describe FSR in the muon channel [5].

The rates of events with large-angle additional ISR are found to be consistent in the muon $(2.83 \pm 0.06)\%$ and kaon $(2.61 \pm 0.08)\%$ data in the 'FSR' subsamples, after the contribution of additional FSR is subtracted from the total rates of events. This is a cross-check of the factorization of additional ISR in the muon and kaon processes and justifies the assumption that the loss of $\chi^2$ efficiency due to additional ISR in the kaon channel can be estimated from the muon data [Eq. (3)].

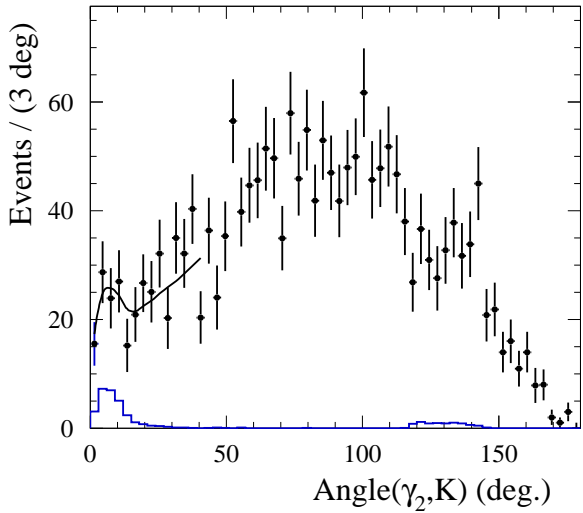

FIG. 10: Angle (degrees) between the additional photon and the closest kaon for data after background subtraction (points) and $KK$ MC (histogram). The plot corresponds to the 'FSR' sample in the [0.95–1.1] GeV/$c^2$ mass region. The MC is normalized to the data luminosity. The fit for the data/MC comparison for the amount of FSR events is also shown (solid line).

The data/MC ratio of efficiencies $C_{\chi^2}^{\mu\mu,\text{FSR sub.}}$ of the tight $\chi^2$ selection for the muons is shown in Fig. 11, where events with additional FSR are subtracted both in data and MC. The bins in the $J/\psi$ and $\psi(2S)$ vicinity are removed, as the different kinematics of the narrow resonance decays, present in data only, might bias the $\chi^2$ efficiency ratio. A conservative systematic error of 1% between 3 and 4 GeV/$c^2$ and 2% beyond 4 GeV/$c^2$ is assigned to account for possible uncertainties in the FSR subtraction at large masses.

### 2. Effects of secondary interactions for kaons

Most effects of secondary interactions are included in the tracking efficiency because of the tight requirements imposed on the track pointing to the interaction region. The minor residual effect on the $\chi^2$-selection efficiency is estimated from simulation, and normalized to the data using the observed rate of interacting kaons in the $KK\gamma$ sample.

Interactions are tagged in the data and MC samples by tracks with transverse impact parameter in the high range $0.15 < \text{doca}_{xy} < 0.5$ cm. According to the simulation, this method identifies about 51% of the events with secondary interactions. The sample is restricted to the intermediate $\chi^2$ region to enhance the interaction rates, while keeping the backgrounds at manageable levels. The background-subtracted distribution (Fig. 12) of the larger $\text{doca}_{xy}$ of the two kaons in the event ($\text{doca}_{xy}^{\text{max}}$) exhibits a striking difference with the corresponding distribution for muons, as expected from secondary interactions. The muon distribution is assumed to describe the contribution of non-interacting kaon tracks, after normalization to the kaon distribution in the region $\text{doca}_{xy}^{\text{max}} < 0.05$ cm; the interacting kaon contribution is taken as the complementary distribution. Using the rates of interacting kaons with $0.15 < \text{doca}_{xy}^{\text{max}} < 0.5$ cm, we find that the simulation underestimates the level of secondary interactions by a factor of $1.51 \pm 0.07 \pm 0.09$, where the first error is statistical and the second is systematic (the systematic uncertainty is derived from the shape difference of the $\text{doca}_{xy}$ distributions in data and MC).

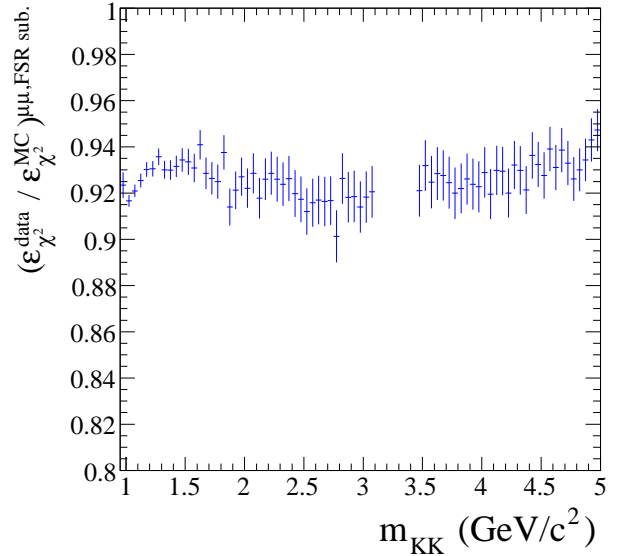

FIG. 11: Data/MC correction for the efficiency of the tight $\chi^2$ selection of $\mu\mu(\gamma)\gamma_{\text{ISR}}$ events, as a function of the $KK$ mass computed using muon-track momenta with assigned kaon mass. The bins in the $J/\psi$ and $\psi(2S)$ vicinity are removed (see text).

The event loss due to interactions in data is extrapolated to the BG region using the normalization factor determined above in the intermediate $\chi^2$ region. A conservative systematic uncertainty of half the loss observed in MC in the BG region is assigned to this extrapolation.

As a test of the contribution of interactions at large $\chi^2$, Fig. 13 shows the $\chi^2_{\text{ISR}}$ distributions for data and MC events at the $\phi$ mass with $0.15 < \text{doca}_{xy}^{\text{max}} < 0.5$ cm. The MC normalization to the data luminosity is corrected for the data/MC ratio of secondary interaction rates. Good agreement is observed over the entire $\chi^2$ range.

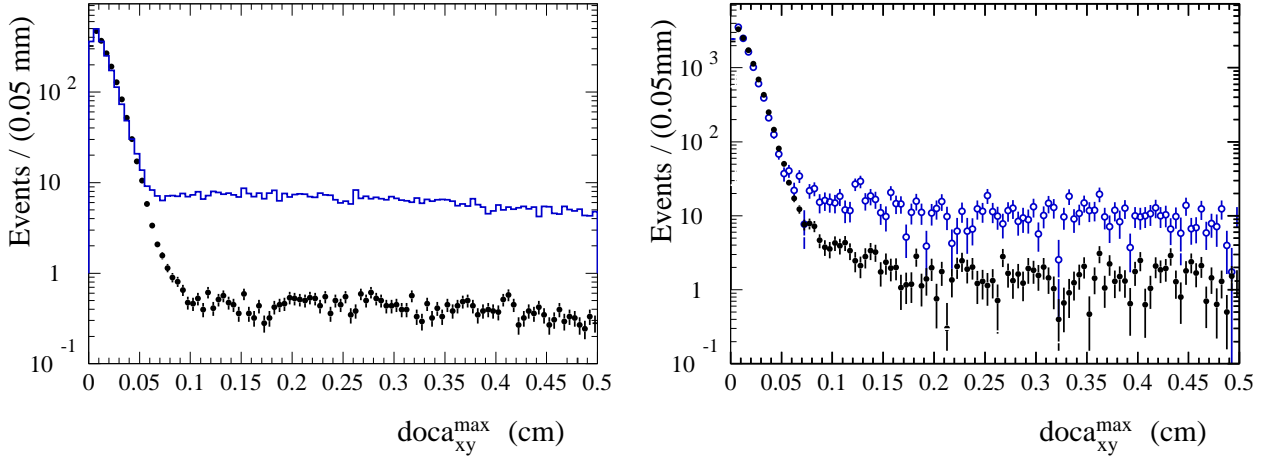

FIG. 12: Distribution of the larger of the two transverse distances of closest approach to the interaction point ($\text{doca}_{xy}^{\max}$), for muons (black points) and kaons (blue histogram left, blue circles right) for MC (left) and data (right) in the intermediate $\chi^2$ region. The $\mu^+\mu^-$ plots are normalized to the $K^+K^-$ results in the region of $\text{doca}_{xy}^{\max} < 0.05\,\text{cm}$.

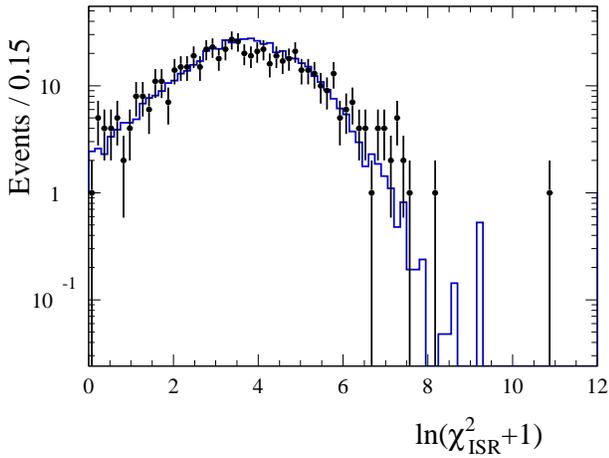

FIG. 13: $\chi^2_{\text{ISR}}$ distribution of $KK\gamma$ events in data (points) and MC (histogram) with $0.98 < m_{KK} < 1.04\,\text{GeV}/c^2$ and $0.15 < \text{doca}_{xy}^{\max} < 0.5\,\text{cm}$. The MC is normalized to the data luminosity and corrected for data/MC differences in secondary interactions.

### 3. Summary of $\chi^2$ efficiency corrections

Figure 14 shows the total data/MC correction $C_{\chi^2}$ for the efficiency of the tight $\chi^2$ selection of the $KK(\gamma)\gamma_{\text{ISR}}$ data. It includes the effects of secondary interactions with the corresponding data/MC correction, and the correction derived from the studies with muons. The total systematic uncertainty on $C_{\chi^2}$ is $2 \times 10^{-3}$ at the $\phi$ mass, slowly increasing to $5 \times 10^{-3}$ at $3\,\text{GeV}/c^2$. Above $3\,\text{GeV}/c^2$, the systematic errors are dominated by the uncertainty of the FSR subtraction.

## IV. BACKGROUND STUDIES

Backgrounds in the $KK(\gamma)\gamma_{\text{ISR}}$ sample stem primarily from other ISR events: $\pi^+\pi^-\gamma$, $\mu^+\mu^-\gamma$, $K^+K^-\eta\gamma$, $K^+K^-\pi^0\gamma$, $\pi^+\pi^-\pi^0\gamma$, $\pi^+\pi^-2\pi^0\gamma$, $p\bar{p}\gamma$, and $K_S K_L\gamma$. These types of events are included in the candidate sample if a (double) mis-ID occurs or if the photons from a $\pi^0$ or $\eta$ decay are not reconstructed. Non-ISR $q\bar{q}$ and $\tau^+\tau^-(\gamma)$ events represent other sources of background. In the latter cases, an energetic photon from $\pi^0$ decay is misidentified as the ISR photon.

Simulated ISR samples are normalized to the luminosity of the data, rescaled to the production cross sections measured with *BABAR* when available [15–18]. Backgrounds from $\mu^+\mu^-\gamma$ and $\pi^+\pi^-\gamma$ events are kinematically confined to the same tight 2D-$\chi^2$ region as the signal, and are separated from the kaon channel by PID only (Sec. III C). The background spectra shown in Fig. 15 are deduced from MC, normalized to the data luminosity, with mis-ID probabilities corrected for data/MC differences. Events from the $\rho \to \pi\pi$ resonance, misidentified as '$KK$' events, peak at $m_{KK} \approx 1.2\,\text{GeV}/c^2$. They represent about 20% of the data at that point and much less everywhere else, while the $\mu\mu \to$ '$KK$' background is a sizeable fraction of the sample only at threshold and at large $m_{KK}$. In addition, since the $J/\psi$ is not included in AfkQed, $0.42 \pm 0.18$ events are subtracted to account for the $J/\psi \to \mu\mu \to$ '$KK$' background, where the uncertainty includes the statistical component and the mis-ID systematic uncertainty. The subtraction is performed in the shifted $m_{KK}$ range $[3.2–3.3]\,\text{GeV}/c^2$.

ISR channels with higher multiplicities populate wide regions of the 2D-$\chi^2$ plane. They are studied with MC in three mass ranges. In the $m_{KK}$ region below $1.1\,\text{GeV}/c^2$, the multibody ISR background is dominated by $K^+K^-\eta\gamma$ events, whose distribution peaks in the $\phi$ resonance region. However, this background, with many additional photons in the final state, is efficiently removed by the $\chi^2$ selection.

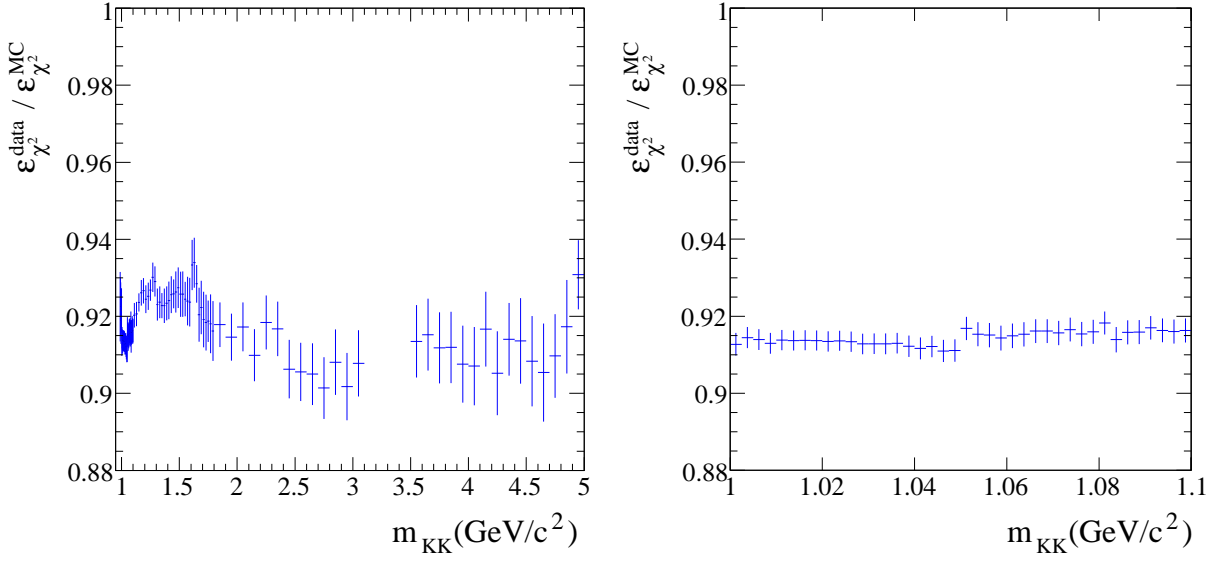

FIG. 14: Total data/MC correction for the efficiency of the tight $\chi^2$ selection as a function of the $KK$ mass. The figure on the right is a zoom of the figure on the left in the $\phi$ resonance region.

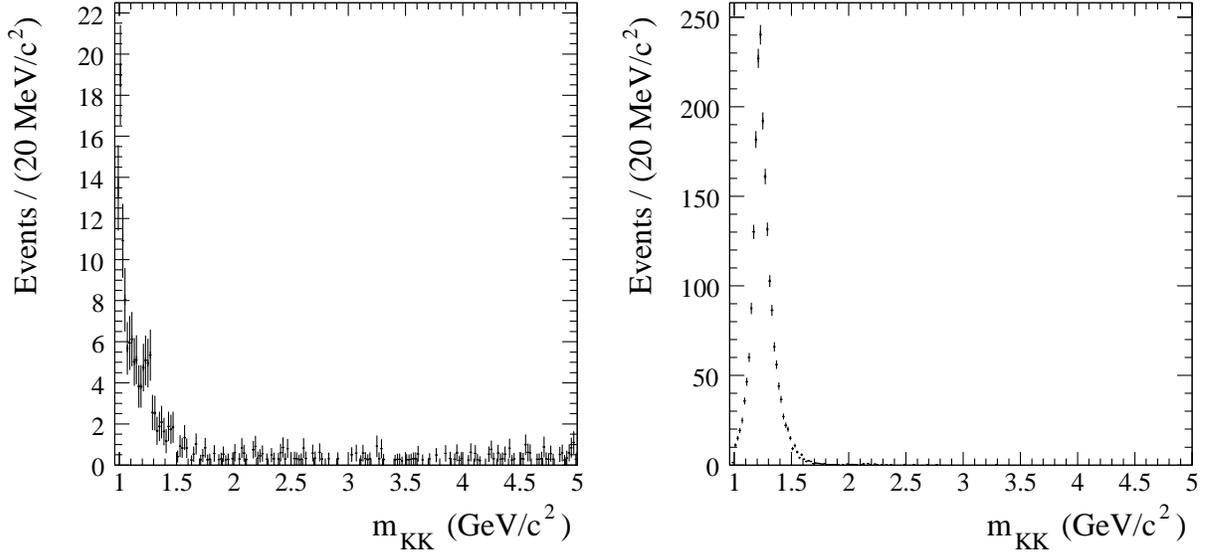

FIG. 15: The $\mu\mu \to$ '$KK$' (left) and $\pi\pi \to$ '$KK$' (right) backgrounds corrected for data/MC differences in mis-ID, as a function of $m_{KK}$.

The JETSET fragmentation model used to generate $q\bar{q}$ MC events might not describe low-multiplicity final states with the required accuracy. The normalization of the $q\bar{q}$ MC sample is consequently performed using data. Background from $q\bar{q}$ events is due to photons from $\pi^0$ decays that are mistaken as the ISR photon candidate, either when the two photons merge in the same EMC cluster, or when the most energetic photon is selected.

In the latter case, the primary $\pi^0$ can be reconstructed by pairing the ISR photon candidate with an additional detected photon. The comparison of the $\pi^0$ yields in data and MC provides the $q\bar{q}$ MC sample normalization.

The $\pi^0$ yields are studied in three $m_{KK}$ intervals: [threshold–1.1], [1.1–3], and [3–5] GeV/$c^2$. To enhance the $\pi^0$ rate significance, a 20 MeV/$c^2$ band centered on the $\phi$ mass is removed from the first interval, and the tight $\chi^2$ region is further reduced to $\ln(\chi^2_{ISR} + 1) > 1$. Normalization factors are determined for all mass ranges in the intermediate $\chi^2$ region and, in the reduced-tight $\chi^2$ region, for masses above the $\phi$ resonance. The normal-ization factor for the tight $\chi^2$ selection cannot be directly assessed from data in the signal-dominated mass range below 1.1 GeV/$c^2$.

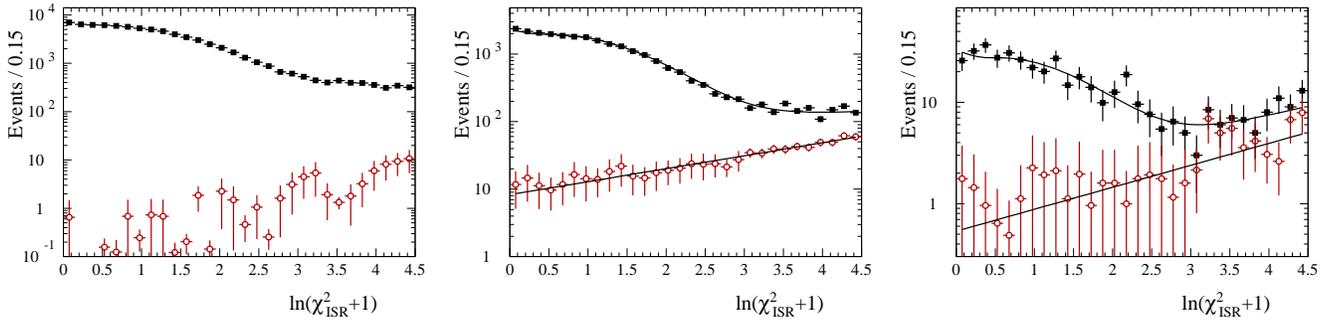

FIG. 16: The data $\ln(\chi^2_{\mathrm{ISR}} + 1)$ distributions (black squares) in the [0.98–1.1] GeV/$c^2$ (left), [1.1–3] GeV/$c^2$ (middle) and [3–5] GeV/$c^2$ (right) $KK$ mass regions, after subtracting the $\pi\pi$ and $\mu\mu$ backgrounds. The (red) open points show the contributions of the remaining backgrounds normalized as described in the text. The solid line represents the result of the fit to the data distributions in the second and third $m_{KK}$ regions, where the signal shape is taken from the [0.98–1.1] GeV/$c^2$ region, and the background shape from MC.

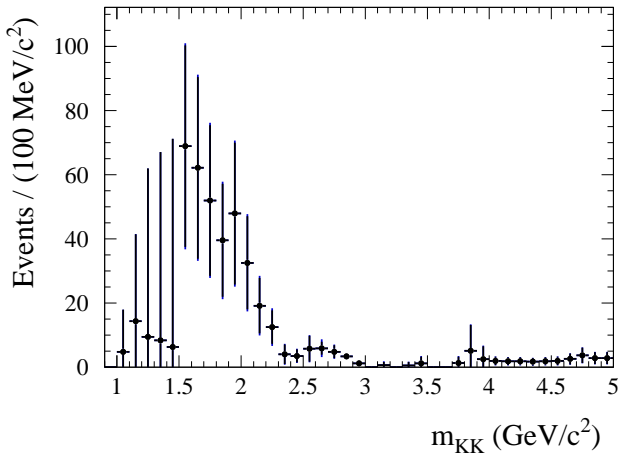

FIG. 17: Total estimated background ($\mu^+\mu^-\gamma$ and $\pi^+\pi^-\gamma$ backgrounds not included) with the tight $\chi^2$ selection. The error bars are dominated by the correlated systematic errors due to the normalization factors of the $q\bar{q}$ MC and the $K^+K^-\eta\gamma$ MC.

While the intermediate $\chi^2$ region is populated by $K^+K^-\pi^0\pi^0$ events, possibly with extra particles, the JETSET simulation indicates that $K^+K^-\pi^0$ is the dominant $q\bar{q}$ background in the tight $\chi^2$ region. It amounts to 73% and 84% of the $q\bar{q}$ background, respectively, in the two highest $m_{KK}$ intervals defined above. Although the background fraction is only at the level of $10^{-3}$ in the $\phi$ mass region, control of the $K^+K^-\pi^0$ component is important, as it is topologically indistinguishable from the $KK(\gamma)_{\mathrm{ISR}}$ signal.

In the intermediate $\chi^2$ region, dominated by the multi-$\pi^0$ backgrounds, the $\pi^0$ yield is extracted from the mea-sured $\gamma_{\mathrm{ISR}}\gamma$ invariant mass distributions. In the tight $\chi^2$ region, dominated by the $K^+K^-\pi^0$ component, the photon momenta are best determined by the (small $\chi^2_{\mathrm{FSR}}$) FSR fit, and the $\pi^0$ signal is extracted from the $\gamma_{\mathrm{ISR}}\gamma$ invariant mass distribution obtained with the fitted momenta. To verify that similar normalization factors apply to the $K^+K^-\pi^0$ and multi-$\pi^0$ components, the backgrounds expected at masses larger than 1.1 GeV/$c^2$ are compared, whether directly estimated in the tight $\chi^2$ region or extrapolated from the intermediate $\chi^2$ region. Although the background composition varies with mass and across the 2D-$\chi^2$ plane, the data-MC $q\bar{q}$ normalization factors obtained in different $\chi^2$ regions are consistent with each other to within the statistical uncertainties in all mass intervals investigated. A conservative systematic error is assigned.

As a test of the normalization procedure using $\pi^0$ tagging, we alternatively deduce the $q\bar{q}$ background normalization factors from a fit of the $\chi^2_{\mathrm{ISR}}$ distributions in data in the [1.1–3] and [3–5] GeV/$c^2$ $m_{KK}$ intervals (Fig. 16). For this test, the $\pi\pi\gamma$ and $\mu\mu\gamma$ backgrounds are subtracted, as obtained from simulation and PID studies. The signal shape of the $\chi^2_{\mathrm{ISR}}$ distribution is taken from the almost background-free data in the [0.98–1.1] GeV/$c^2$ mass region. The background shape is from MC. In contrast with the FSR fit, for which the dominant $K^+K^-\pi^0$ background component returns a good $\chi^2_{\mathrm{FSR}}$, the background presents a $\chi^2_{\mathrm{ISR}}$ distribution shifted to high values with respect to signal. The fitted background contributions obtained in the two high-mass intervals are consistent within errors with the JETSET expectation scaled with the normalization factors deduced from the $\pi^0$ yields. This test thus validates the normalization procedure and confirms that the remaining backgrounds (e.g., a non-$\pi^0$ component) are within the quoted systematic uncertainties.

The total background after the tight $\chi^2$ selection is shown in Fig. 17. The distribution does not include the $\mu^+\mu^-\gamma$ and $\pi^+\pi^-\gamma$ contributions shown separately in Fig. 15. In the $m_{KK}$ region below $1.1\,\text{GeV}/c^2$, the non-ISR background fraction is at the $10^{-3}$ level, with a conservative uncertainty assigned for the normalization factor.

## V. UNFOLDING OF THE MASS SPECTRUM

The distribution of $KK(\gamma)\gamma_{\text{ISR}}$ events as a function of $\sqrt{s'}$ is deduced from the background-subtracted $m_{KK}$ spectrum through unfolding. Prior to unfolding, the mass spectrum is corrected for data/MC efficiency differences [Eq. (2)]. As the level of additional FSR is very small for kaons, $m_{KK}$ differs from $\sqrt{s'}$ only through resolution spreading. Because the $\phi$ resonance is narrow, accurate unfolding is critical to obtain the true line shape. Resolution uncertainties affect the unfolded $\phi$ width; however the iterative unfolding method used in this analysis, as described below, is mostly insensitive to a precise mass calibration and differences between the physics (unfolded) spectra in MC and data.

### A. Mass calibration and resolution studies

Mass calibration and mass resolution tests are provided through a study of $K_S^0 \to \pi^+\pi^-$ decays, from a sample of ISR-produced $\phi$ mesons decaying into $K_S^0 K_L^0$. The resolution measured with data is compared to MC, while the reconstructed $K_S^0$ mass is compared to MC results and the nominal value [19]. In the $\phi \to K^+K^-$ decays, the mass calibration and resolution are governed by the measurement of the opening angle, because the $\phi$ mass lies very close to the $K^+K^-$ threshold. In contrast, the momentum measurement controls the mass measurement of the $J/\psi \to \mu\mu$ decays, which provide a calibration of the momentum scale. In the $K_S^0 \to \pi^+\pi^-$ decays, both angular and momentum measurements contribute to the resolution, but the momentum measurement plays a minor role as in $\phi \to K^+K^-$ decays. As a result, the $K_S^0$ sample is particularly relevant to the understanding of the $\phi$ resonance parameters.

No significant shift is observed in data in the $K_S^0$ sample between the reconstructed mass and the nominal one [19]. After correction for the different mean values of track momentum and opening angle in $K_S^0$ and $\phi$ decays, the $\phi$ mass shift is found to be consistent with zero. A conservative systematic uncertainty on the $\phi$ mass scale of $0.052\,\text{MeV}/c^2$ is assigned, dominated by the limited number of events in the $K_S^0$ sample.

A few-percent difference is observed between mass resolutions in the data and MC $K_S^0$ samples. After correction for the mean momentum and opening angle in $\phi$ decays, this translates into a bias on the $\phi$ width after unfolding of $\Delta\Gamma_\phi = 0.020\pm0.043\,\text{MeV}$. As for the calibration, no correction is applied. A systematic uncertainty of $0.063\,\text{MeV}$ is assigned to the fitted $\phi$ width.

### B. Unfolding procedure

This analysis follows the same iterative unfolding procedure as used for the pion cross section analysis, described in detail in Ref. [5].

When starting the unfolding procedure, significant differences are observed between the reconstructed mass spectra in data and MC, close to threshold, as well as at large masses (Fig. 18). To minimize biases, the unfolding is performed iteratively. The transfer matrix (Fig. 19), initially taken from MC, is improved at each step, to bring the shape of the reconstructed MC mass spectrum into better agreement with the data.

The first unfolding step corrects the main resolution effects on the data spectrum (Fig. 20). The result is compared to the physics MC spectrum and used to improve the transfer matrix through reweighting of the latter. After reweighting, almost all systematic differences between data and reconstructed MC are removed, and further iterations do not improve the result. The effect of the second iteration is used to estimate the systematic uncertainty, in addition to a closure test using known distributions (close to data) in a large set of pseudo-experiments. The overall unfolding correction on the $KK(\gamma)$ cross section at the $\phi$ peak amounts to about 15%, as seen in Fig. 20.

## VI. ACCEPTANCE CORRECTIONS

The overall acceptance $\varepsilon_{KK\gamma}$ entering Eq. (1) is calculated using the AfkQed generator and full simulation of the $KK(\gamma)\gamma_{\text{ISR}}$ events. The overall acceptance $\varepsilon_{\mu\mu\gamma}$, which enters the effective luminosity calculation (Sec. VII A), is estimated in the same way for the $\mu\mu(\gamma)\gamma_{\text{ISR}}$ events. Both $\varepsilon_{KK\gamma}$ and $\varepsilon_{\mu\mu\gamma}$ are corrected for differences between data and simulation. Corrections for differences in efficiencies for detector simulation (Sec. III) are applied prior to unfolding, independently for each channel. This section deals with geometrical acceptance corrections, which apply to the $KK(\gamma)\gamma_{\text{ISR}}/\mu\mu(\gamma)\gamma_{\text{ISR}}$ ratio.

Given the small fraction of FSR for kaons, the additional FSR generation with PHOTOS is found to agree with data to an adequate precision (Sec. III D 1). The FSR prescription is also found to be in agreement with data for the muons [5]. In contrast, the additional ISR generation by AfkQed leads to large discrepancies with data as studied in detail in Sec. III D and Ref. [5]. Additional ISR issues are, however, common to the kaon and muon channels, and corrections to the geometrical acceptance cancel in the $KK(\gamma)\gamma_{\text{ISR}}/\mu\mu(\gamma)\gamma_{\text{ISR}}$ ratio to first order. Second-order corrections are induced by the different kinematic conditions in the two channels. Kinematic effects of the approximate NLO ISR on the acceptance, including related effects on the 'preselection cut' (defined in Sec. II), are studied at the four-vector level with large samples of events generated with AfkQed

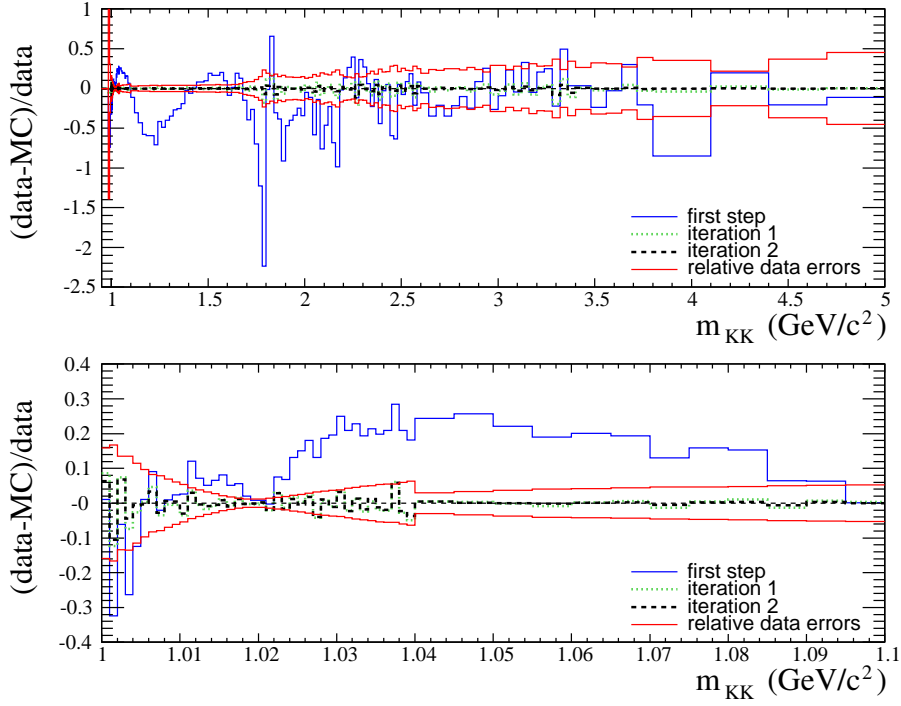

FIG. 18: (color online). Relative difference between data and reconstructed MC in the tight $\chi^2$ region, at the first step (blue histogram), after one iteration (dotted green line) and after a second iteration (dashed black line). The diagonal elements of the error matrix are indicated by the red histograms. The bottom plot is a zoom of the top plot in the [1–1.1] GeV/$c^2$ $m_{KK}$ range.

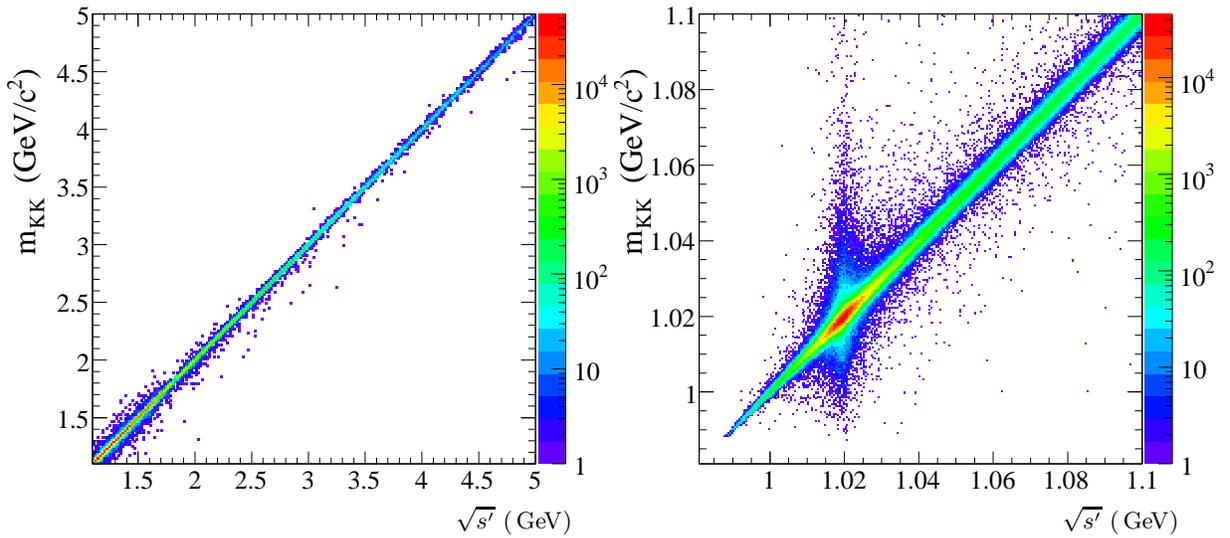

FIG. 19: (color online). Transfer matrix for events in the tight $\chi^2$ region. The plot on the right is a zoom of the left plot in the [1–1.1] GeV/$c^2$ $m_{KK}$ range.

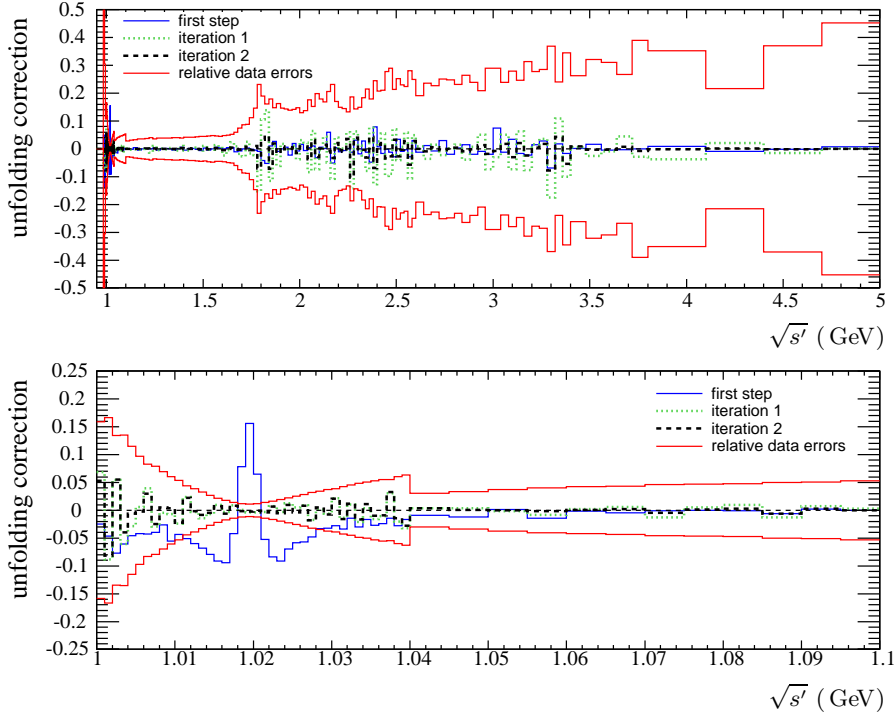

FIG. 20: (color online). Relative correction applied to the data spectrum at the first step of the unfolding (blue histogram), after one iteration (dotted green line) and after a second iteration (dashed black line). The diagonal elements of the error matrix are indicated by the red histograms. The bottom plot is a zoom of the top plot in the $[1-1.1]\,\mathrm{GeV}/c^2\ m_{KK}$ range.

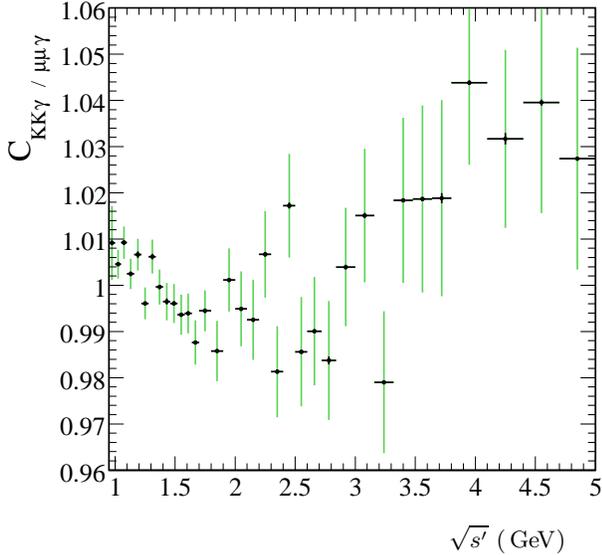

FIG. 21: (color online). The full correction to the $KK\gamma/\mu\mu\gamma$ acceptance ratio to account for data/MC differences for additional ISR and secondary interaction effects. The vertical black error bars show the small but fully correlated errors coming from the data/MC correction of secondary interactions. The green error bars show the total error in each bin.

and Phokhara. The ratio of acceptances in $KK\gamma$ is com-

pared to the corresponding ratio in $\mu\mu\gamma$ in the same mass range. The correction on the double ratio

$$C_1 = \left(\frac{\varepsilon_{KK\gamma}^{\mathrm{Phokhara}}}{\varepsilon_{KK\gamma}^{\mathrm{AfkQed}}}\right)_{\mathrm{acc+presel}}^{\mathrm{gen}} \Bigg/ \left(\frac{\varepsilon_{\mu\mu\gamma}^{\mathrm{Phokhara}}}{\varepsilon_{\mu\mu\gamma}^{\mathrm{AfkQed}}}\right)_{\mathrm{acc+presel}}^{\mathrm{gen}} \quad (4)$$

is very much reduced with respect to corrections for each channel, due to the cancelation of generator effects in the ratio. The correction of a few per mil in the $\phi$ region increases to 1-2 percent in the $[1.5-4]\,\mathrm{GeV}/c^2$ interval and to 3-4 percent (with larger errors) at higher masses.

In addition, the 'preselection cut' efficiency is affected by secondary interactions of kaons in the detector. Estimation of this kaon-specific contribution is studied with full simulation. Interaction effects are inferred from the ratio of the 'preselection cut' efficiencies in the $KK\gamma$ and $\mu\mu\gamma$ full simulation, with a correction to account for the different kinematics. The latter is taken as the ratio of efficiencies at the generator level. The double ratio

$$C_2 = \left(\frac{\varepsilon_{KK\gamma}^{\mathrm{AfkQed}}}{\varepsilon_{\mu\mu\gamma}^{\mathrm{AfkQed}}}\right)_{\mathrm{presel}}^{\mathrm{full}} \Bigg/ \left(\frac{\varepsilon_{KK\gamma}^{\mathrm{AfkQed}}}{\varepsilon_{\mu\mu\gamma}^{\mathrm{AfkQed}}}\right)_{\mathrm{presel}}^{\mathrm{gen}} \quad (5)$$

is at the level of a few per mil. The contribution of secondary interactions to the $KK\gamma/\mu\mu\gamma$ acceptance ratio is scaled in data by the measured data/MC rate of interactions, $1.51 \pm 0.11$ (Sec. III D 2).

Kinematic effects on the ISR photon efficiency are found to induce a negligible correction to the $KK\gamma/\mu\mu\gamma$

ratio; a systematic error of $1.2 \times 10^{-3}$ is assigned to account for the different sampling of the ISR photon efficiency map.

The overall correction

$$C_{KK\gamma/\mu\mu\gamma} = C_1 \left[ 1 + (1.51 \pm 0.11) \ (C_2 - 1) \right] , \qquad (6)$$

to be applied to the $\varepsilon_{KK\gamma}/\varepsilon_{\mu\mu\gamma}$ acceptance ratio is shown as a function of mass in Fig. 21. The full correction is found to be considerably smaller (and better known) than the precision on the measurement of the $KK\gamma$ spectrum itself. The systematic error displayed in Fig. 21 includes the uncertainty on the ISR photon efficiency and the uncertainty on the data/MC ratio of secondary interaction rates.

## VII.   RESULTS

### A.   The effective ISR luminosity

The effective ISR luminosity is obtained directly from the analysis of $\mu\mu(\gamma)\gamma_{ISR}$ events with the same data, with methods described in detail in Ref. [5]. The effective ISR luminosity $dL_{ISR}^{eff}/d\sqrt{s'}$ is related to the $\sqrt{s'}$ spectrum of $\mu\mu(\gamma)\gamma_{ISR}$ events by

$$\frac{dN_{\mu\mu(\gamma)\gamma_{ISR}}}{d\sqrt{s'}} \ = \ \frac{dL_{ISR}^{eff}}{d\sqrt{s'}} \ \varepsilon_{\mu\mu\gamma}(\sqrt{s'}) \ \sigma_{\mu\mu(\gamma)}^{0}(\sqrt{s'}) \ (1 + \delta_{FSR}^{\mu\mu}(\sqrt{s'})) , \qquad (7)$$

where $dN_{\mu\mu(\gamma)\gamma_{ISR}}/d\sqrt{s'}$ is obtained by unfolding the observed $m_{\mu\mu}$ distribution, $\varepsilon_{\mu\mu\gamma}$ is the full acceptance for the event sample, determined using MC with corrections from data, $\delta_{FSR}^{\mu\mu} = |FSR|^2/|ISR|^2$ accounts for the leading order (LO) FSR contribution to the $\mu\mu\gamma$ final state, and $\sigma_{\mu\mu(\gamma)}^{0}$ is the bare cross section calculated with QED for the process $e^+e^- \to \mu^+\mu^-(\gamma)$ (including additional FSR). The LO FSR correction $\delta_{FSR}^{\mu\mu}$ is evaluated using AfkQed at the generator level. The luminosity $dL_{ISR}^{eff}/d\sqrt{s'}$ thus defined integrates over all configurations with up to two ISR photons where at least one photon has $E_\gamma^* > 3$ GeV and $20° < \theta_\gamma^* < 160°$. It includes vacuum polarization, so that the bare $K^+K^-(\gamma)$ cross section is obtained when inserting this effective luminosity into Eq. (1).

The effective ISR luminosity as a function of $\sqrt{s'}$ is determined in 50 MeV bins, which is insufficient near narrow resonances ($\omega$ and $\phi$) because of the rapid variation of the hadronic vacuum polarization term. Therefore, in each 50 MeV bin, we take the local variation from the product of the LO QED luminosity function [3, 4] and the VP factor, and normalize the result to the effective luminosity determined in that bin. In this way, the detailed local features of the vacuum polarization are incorporated, while preserving the measured effective luminosity as a function of mass. To minimize the bin-to-bin statistical fluctuations, the distribution in 50 MeV bins is smoothed before the VP correction is applied, by averaging five consecutive bins (sliding bins). The reduced local error is compensated by the correlation between neighbouring bins. No VP correction is applied for the $J/\psi$ and $\psi(2S)$, as this correction would affect the structure of the resonances themselves.

The statistical errors on the ISR effective luminosity from the measurement of efficiencies are included in the statistical covariance matrix, while the systematic uncertainties from the different corrections are accounted for separately. These uncertainties are $0.3 \times 10^{-3}$ for trigger, $1.3 \times 10^{-3}$ for tracking, $3.2 \times 10^{-3}$ for $\mu$-ID, including the uncertainty on the correlated loss of $\mu$-ID for both tracks, and $1.0 \times 10^{-3}$ for acceptance. The total systematic error on the ISR luminosity amounts to $3.7 \times 10^{-3}$. It is conservatively increased in the [3–5] GeV interval (up to $1-2\%$) to account for the fact that the QED test [5] is performed only at lower masses, and for the increase of the LO FSR correction $\delta_{FSR}^{\mu\mu}$. In addition, a systematic uncertainty of $2.5 \times 10^{-3}$ is assigned for the VP correction in the $\phi$ region, resulting from the uncertainty on the $\phi$ parameters [19] used in the VP calculation.

### B.   $K^+K^-(\gamma)$ bare cross section

The $K^+K^-(\gamma)$ bare cross section $\sigma_{KK(\gamma)}^{0}(\sqrt{s'})$ (including FSR) is computed according to Eq. (1) from the unfolded spectrum. Background subtraction and corrections for data/MC differences in detector simulation are applied to the mass spectrum prior to unfolding. The global acceptance $\varepsilon_{KK\gamma}$ is obtained with AfkQed (Fig. 22), and corrected by the $C_{KK\gamma/\mu\mu\gamma}$ factor defined by Eq. (6). The effective ISR luminosity $dL_{ISR}^{eff}/d\sqrt{s'}$ is obtained from muon data as explained above.

The $\sigma_{KK(\gamma)}^{0}(\sqrt{s'})$ cross section is shown in Fig. 23, from $K^+K^-$ production threshold up to 5 GeV. Files containing the cross section data and their covariance matrices are provided in the EPAPS repository [20].

The cross section spans more than six orders of magnitude and is dominated by the $\phi$ resonance close to threshold. Other structures are clearly visible at higher masses. The contributions to the $K^+K^-$ final state from the decays of the narrow $J/\psi$ and $\psi(2S)$ resonances have been subtracted for the cross section measurement and for the determination and parametrization of the kaon form-factor (Sec. VII C). The important correlations between the $J/\psi$ ($\psi(2S)$) bin and the neighboring ones, resulting from the subtraction procedure, are taken into account in the covariance matrix. The $J/\psi$ and $\psi(2S)$ branching fractions to $K^+K^-$ are considered separately in Section VII G.

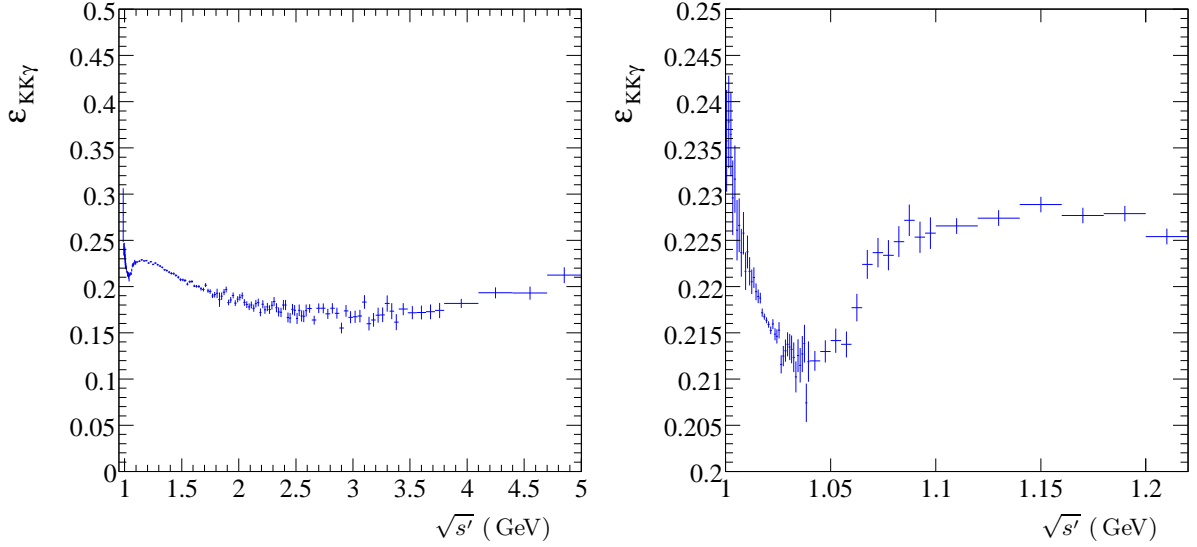

FIG. 22: Global acceptance $\varepsilon_{KK\gamma}$ computed with AfkQed. A zoom on the low mass region is shown in the right plot.

TABLE I: Systematic uncertainties (in units of $10^{-3}$) on the bare cross section for $e^+e^- \to KK(\gamma_{\mathrm{FSR}})$ from the determination of the various efficiencies in different $KK$ mass ranges (in GeV). The statistical part of the efficiency measurements is included in the total statistical error in each mass bin. The last line gives the total systematic uncertainty on the $KK$ cross section, including the systematic error from the ISR luminosity from muons.

| $m_{KK}$ range (GeV) | 0.98-0.99 | 0.99-1 | 1-1.01 | 1.01-1.03 | 1.03-1.04 | 1.04-1.05 | 1.05-1.1 |
|---|---|---|---|---|---|---|---|
| Trigger/ filter | 1.0 | 0.7 | 0.7 | 0.7 | 0.7 | 0.8 | 0.8 |
| Tracking | 1.8 | 1.8 | 1.9 | 2.8 | 2.8 | 2.8 | 5.3 |
| $K$-ID | 10.6 | 8.8 | 5.4 | 4.1 | 6.5 | 12.7 | 12.8 |
| Background | 157.2 | 20.9 | 1.6 | 0.1 | 0.3 | 0.6 | 1.1 |
| Acceptance | 1.6 | 1.6 | 1.6 | 1.6 | 1.6 | 1.6 | 1.6 |
| Kinematic fit ($\chi^2$) | 2.0 | 2.0 | 2.0 | 2.0 | 2.0 | 3.3 | 3.2 |
| ISR luminosity | 3.7 | 3.7 | 3.7 | 3.7 | 3.7 | 3.7 | 3.7 |
| Unfolding | 3.2 | 3.2 | 3.2 | - | 1.2 | 1.2 | 1.2 |
| VP correction | - | - | 0.4 | 2.5 | 0.5 | - | - |
| Sum (cross section) | 157.7 | 23.4 | 8.2 | 7.2 | 8.5 | 14.1 | 14.9 |

| $m_{KK}$ range (GeV) | 1.1-1.2 | 1.2-1.3 | 1.3-1.5 | 1.5-1.7 | 1.7-2.3 | 2.3-3 | 3-4 | 4-5 |
|---|---|---|---|---|---|---|---|---|
| Trigger/ filter | 0.6 | 0.5 | 0.4 | 0.4 | 0.4 | 0.4 | 0.5 | 0.5 |
| Tracking | 7.2 | 8.2 | 8.8 | 9.2 | 9.7 | 10.0 | 10.2 | 10.2 |
| $K$-ID | 13.0 | 16.3 | 26.3 | 33.1 | 41.1 | 51.4 | 52.1 | 54.4 |
| Background | 4.9 | 11.8 | 18.5 | 13.6 | 56.0 | 24.3 | 67.6 | 243.5 |
| Acceptance | 1.6 | 1.6 | 1.6 | 1.6 | 1.6 | 1.6 | 1.6 | 1.6 |
| kinematic fit ($\chi^2$) | 2.3 | 2.5 | 2.6 | 3.5 | 4.5 | 5.6 | 14.6 | 23.4 |
| ISR luminosity | 3.7 | 3.7 | 3.7 | 3.7 | 3.7 | 3.7 | 12.7 | 22.3 |
| Unfolding | 0.7 | 0.7 | 0.7 | - | - | - | - | - |
| VP correction | - | - | - | - | - | - | - | - |
| Sum (cross section) | 16.4 | 22.3 | 33.7 | 37.3 | 70.4 | 58.1 | 88.1 | 251.8 |

Figure 24 shows three enlargements in the [1–2.1] GeV

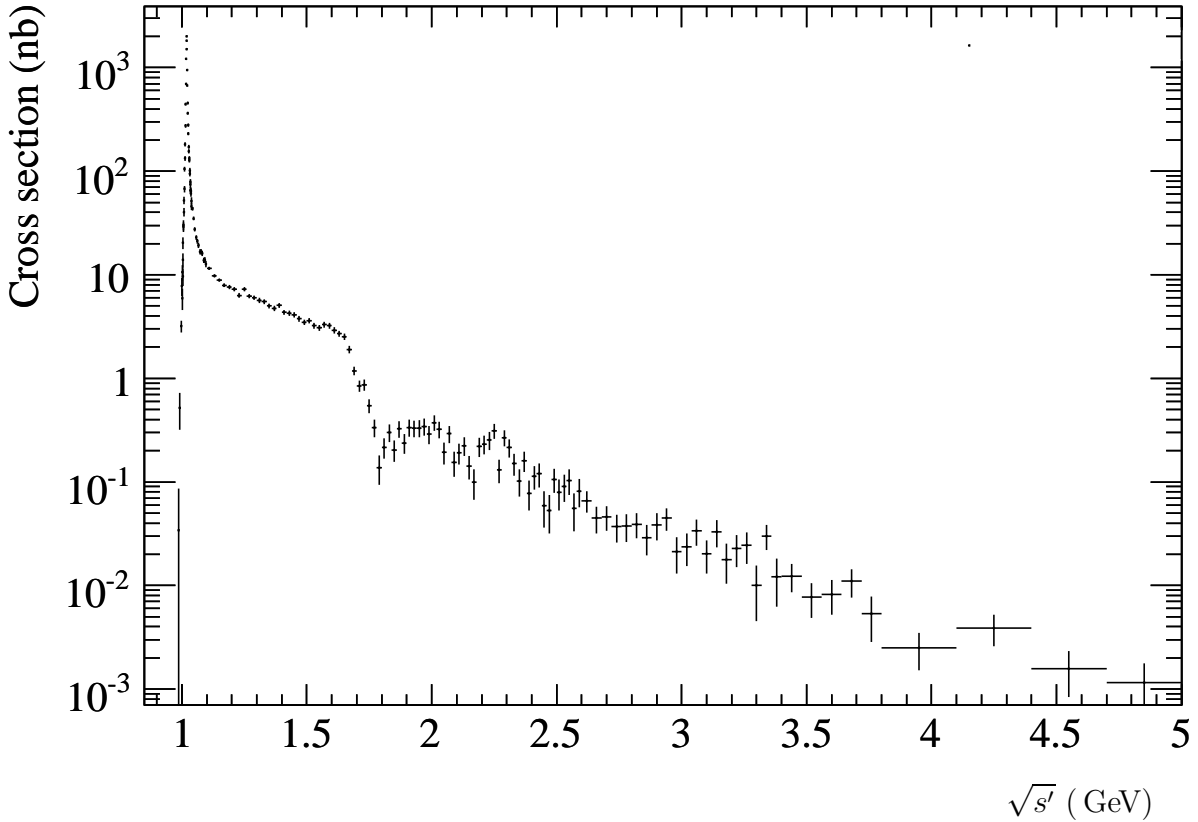

FIG. 23: The measured $e^+e^- \to K^+K^-(\gamma)$ bare cross section (including FSR). Systematic and statistical uncertainties are shown, i.e., the diagonal elements of the total covariance matrix. The contributions of the decays of the $J/\psi$ and $\psi(2S)$ resonances to $K^+K^-$ have been subtracted.

energy interval. Data from previous measurements are also shown. The *BABAR* measurement agrees with the previous results [21–26], but the *BABAR* data cover the full energy range, and are more precise. In particular, the dip around 1.8 GeV is mapped with much increased precision.

The systematic uncertainties affecting the bare $K^+K^-(\gamma)$ cross section are summarized in Table I. The overall systematic uncertainty is $7.2 \times 10^{-3}$ in the [1.01–1.03] GeV mass range, but significantly larger outside the $\phi$ region. All the correlations from the various corrections are fully propagated to the final covariance matrix of the cross section. Each systematic error is treated as fully correlated in all mass bins, except for the ones from the unfolding and the vacuum polarization correction on the luminosity (Sec. VII A). The calibration and resolution uncertainties also affect the final cross section. They exhibit a rapid variation in the $\phi$ region (Fig. 25) as well as strong bin-to-bin anticorrelations (hence they have a negligible effect on the dispersion integral entering the $a_\mu$ calculation). The error on the vacuum polarization correction, which also has important anticorrelations, contributes to the cross section uncertainty, but does not affect the dressed form factor and only slightly the dispersion integral (Sec. VII H).

### C. Charged kaon form factor

The square of the kaon form factor is defined by the ratio of the dressed cross section without final-state interactions, to the lowest-order cross section for point-like spin 0 charged particles

$$|F_K|^2(s') = \frac{3s'}{\pi \alpha^2(0)\beta_K^3} \frac{\sigma_{KK}(s')}{C_{\rm FS}} , \qquad (8)$$

where

$$\sigma_{KK}(s') = \sigma_{KK(\gamma)}^0(s') \ \left(\frac{\alpha(s')}{\alpha(0)}\right)^2 \qquad (9)$$

is the dressed cross section, deduced from the bare cross section $\sigma_{KK(\gamma)}^0$ measured above, $\beta_K = \sqrt{1 - 4m_K^2/s'}$ is the kaon velocity, and $C_{\rm FS} = 1 + \frac{\alpha}{\pi}\eta_K(s')$ is the final-state correction [27–29]. At the $\phi$ mass, the 4.2% deviation from unity of $C_{\rm FS}$ is completely dominated by the Coulomb interaction between $K^+$ and $K^-$. It is slowly decreasing at higher masses. The form factor values and their covariance matrices are provided in the EPAPS repository [20].

For purposes of measuring the $\phi$ resonance parameters and providing an empirical parametrization of the form factor over the full range of the measurement, we fit the kaon form factor with a model [30] based on a sum of resonances. While the parametrized form factor is conveniently compared with the results of experiments at fixed energy values, the fit is necessary to extract the $\phi$ resonance parameters in the presence of other small contributions that need to be determined. Both isospin I = 0 and I = 1 resonances are considered since $KK$ is not an eigenstate of isospin. We express the form factor as:

$$
\begin{aligned}
F_K(s) \;=\; & \left(a_\phi \; BW_\phi + a_{\phi'} \; BW_{\phi'} + a_{\phi''} \; BW_{\phi''}\right)/3 \\
& + \left(a_\rho \; BW_\rho + a_{\rho'} \; BW_{\rho'} + a_{\rho''} \; BW_{\rho''} + a_{\rho'''} \; BW_{\rho'''}\right)/2 \\
& + \left(a_\omega \; BW_\omega + a_{\omega'} \; BW_{\omega'} + a_{\omega''} \; BW_{\omega''} + a_{\omega'''} \; BW_{\omega'''}\right)/6 \;,
\end{aligned}
\tag{10}
$$

with the constraints

$$
\begin{aligned}
a_\phi + a_{\phi'} + a_{\phi''} &= 1, \\
a_\rho + a_{\rho'} + a_{\rho''} + a_{\rho'''} &= 1, \\
a_\omega + a_{\omega'} + a_{\omega''} + a_{\omega'''} &= 1.
\end{aligned}
\tag{11}
$$

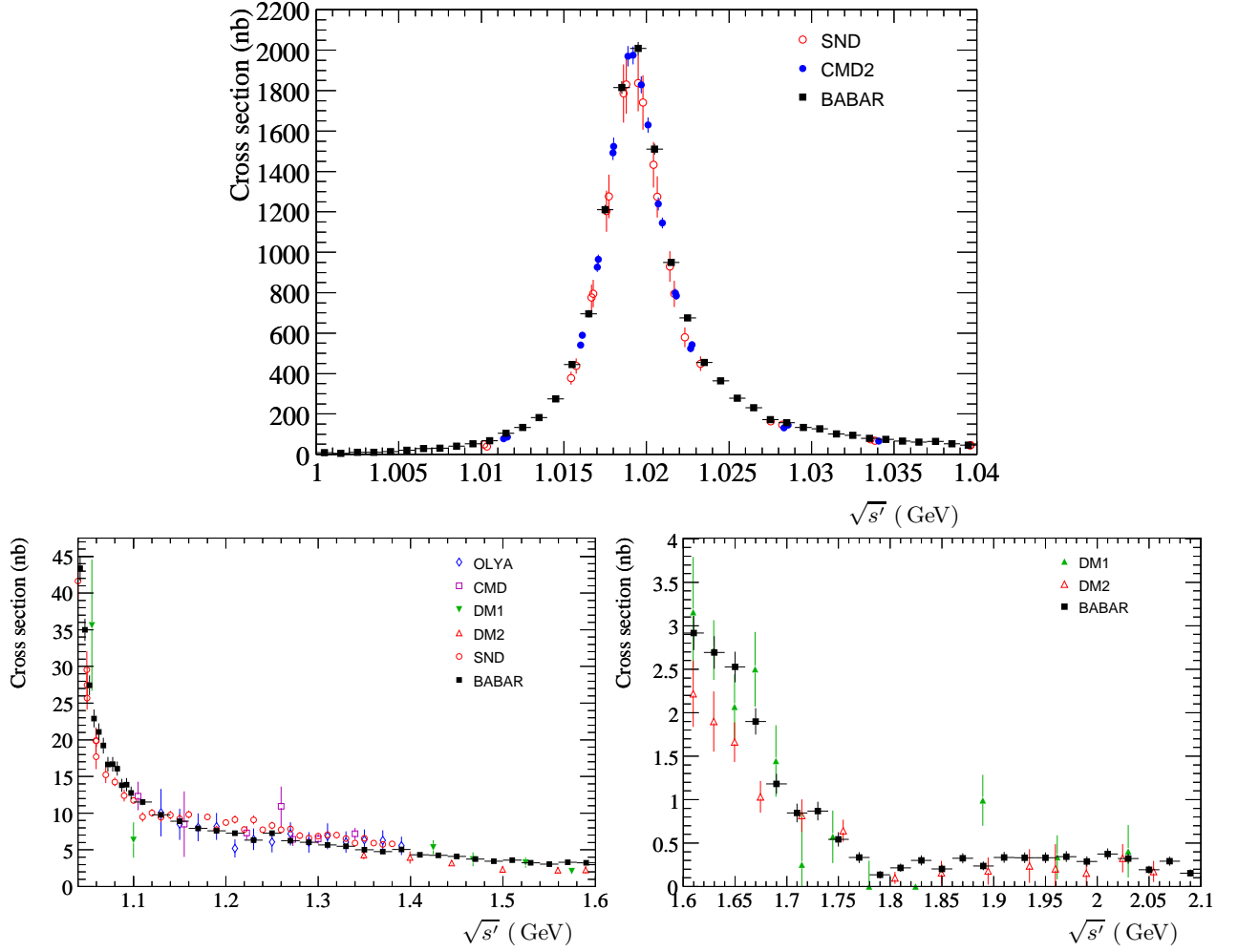

FIG. 24: (color online). The measured $e^+e^- \to K^+K^-$ bare cross section in the [1–1.04] GeV (top), [1.04–1.6] GeV (bottom left), and [1.6–2.1] GeV (bottom right) mass intervals, together with results published by previous experiments. Systematic and statistical uncertainties are shown, i.e., the diagonal elements of the total covariance matrices.

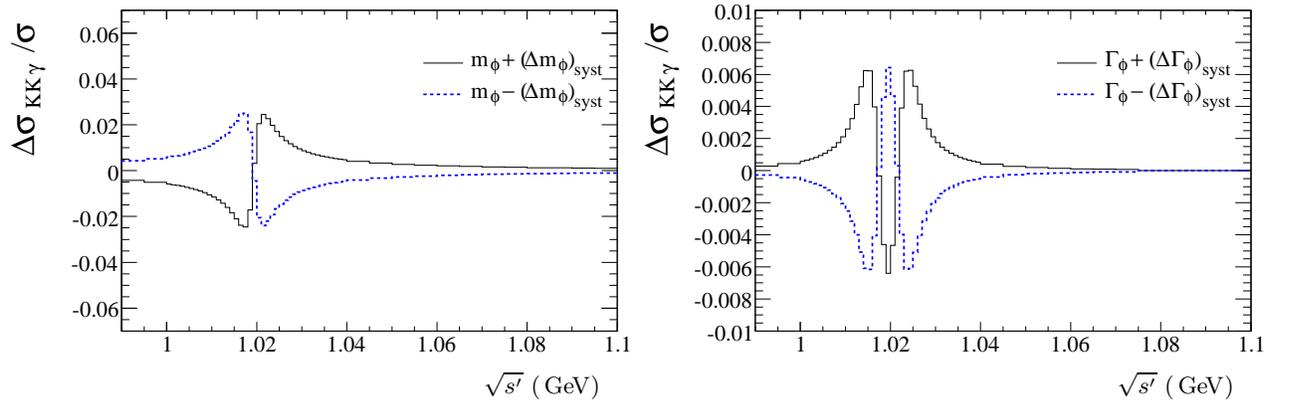

FIG. 25: Relative variations of the $KK(\gamma)$ cross section due to mass calibration (left) and resolution (right) uncertainties. The solid black (dashed blue) histogram indicates the effect corresponding to a $+1$ ($-1$) standard deviation variation of the given parameter.

All the $a_r$ amplitudes are assumed to be real. The resonance shapes are described by Breit-Wigner expressions:

$$
BW(s, m, \Gamma) = \frac{m^2}{m^2 - s - i \; m\Gamma(s)},
\tag{12}
$$

where the width is, in general, energy dependent. For the $\rho$, we use the Kuhn-Santamaria model, where the dependence is given by:

$$
\Gamma_\rho(s) = \Gamma_\rho \frac{s}{m_\rho^2} \left(\frac{\beta(s, m_\pi)}{\beta(m_\rho^2, m_\pi)}\right)^3,
\tag{13}
$$

with $\beta(s, m) = \sqrt{1 - 4m^2/s}$. For the $\phi$, there are separate contributions from different decay modes (with branching fractions $\mathcal{B}$), approximated as

$$
\Gamma_\phi(s) \;=\; \Gamma_\phi \left[ \mathcal{B}(\phi \to K^+K^-) \frac{\Gamma_{\phi \to K^+K^-}(s, m_\phi, \Gamma_\phi)}{\Gamma_{\phi \to K^+K^-}(m_\phi^2, m_\phi, \Gamma_\phi)} + \mathcal{B}(\phi \to K^0\bar{K}^0) \frac{\Gamma_{\phi \to K^0\bar{K}^0}(s, m_\phi, \Gamma_\phi)}{\Gamma_{\phi \to K^0\bar{K}^0}(m_\phi^2, m_\phi, \Gamma_\phi)} \right.
$$
$$
\left. + 1 - \mathcal{B}(\phi \to K^+K^-) - \mathcal{B}(\phi \to K^0\bar{K}^0) \right] \;,
\tag{14}
$$

where $\Gamma_{\phi \to K\bar{K}}(s, m_\phi, \Gamma_\phi)$ is given by Eq. (13) with suitable replacements. A fixed width is used for the $\phi$ decay modes other than $K^+K^-$ and $K^0\bar{K}^0$, as well as for resonances other than $\phi$ and $\rho$.

Known resonances contribute above the $\phi$: isovector ($\rho'$, $\rho''$) and isoscalar ($\omega'$, $\omega''$, $\phi'$) states. Additional resonances ($\rho'''$, $\omega'''$, $\phi''$) are needed in order to fit the structures seen between 1.8 and 2.4 GeV. All the contributions cannot be determined from the charged kaon form factor fit alone. A complete analysis would require the simultaneous fit of the charged and neutral kaon form factors, together with the pion form factor and resonance parameters extracted from inelastic channels such as $4\pi$ and $K\bar{K}\pi$. Such an analysis is beyond the scope of this study. The mass and width of states above the $\phi$ are thus fixed to the world average values [19], while the respective amplitudes are fitted.

According to a well-known effect [31], the $\chi^2$ minimization returns fitted values that are systematically shifted with respect to the data points when the full covariance matrix is used in the fit. This feature is due to the correlations, which here arise from both statistical and systematic origins, mostly from the ISR luminosity 50 MeV sliding bins, and systematic errors. To circumvent the problem, we fit the data with only diagonal errors to obtain the central values of the fitted parameters. The error on each parameter is taken as the largest error obtained from the fit either with the full covariance matrix or with only diagonal errors.

The 17-parameter phenomenological fit provides a fair description of $BABAR$ data (Fig. 26) from threshold up to 2.4 GeV ($\chi^2/\text{DoF} = 141.1/100$). The partial $\chi^2$ in the $\phi$ resonance region ([1–1.1] GeV) accounts for 54.4 units, for 52 fitted points. A more accurate comparison is given in Fig. 27, which shows the relative difference between the charged kaon squared form factor from the $BABAR$ data and the fit. While the agreement is in general very good, some oscillations are observed at 1.25 and 1.7 GeV. They correspond to regions where the differences between the data and MC spectra, at the beginning of the unfolding procedure, are relatively large (Fig. 18). While the unfolding correction is almost negligible for the oscillation at 1.25 GeV, one iteration slightly enhances the oscillation at 1.7 GeV, which is thus probably a real effect.

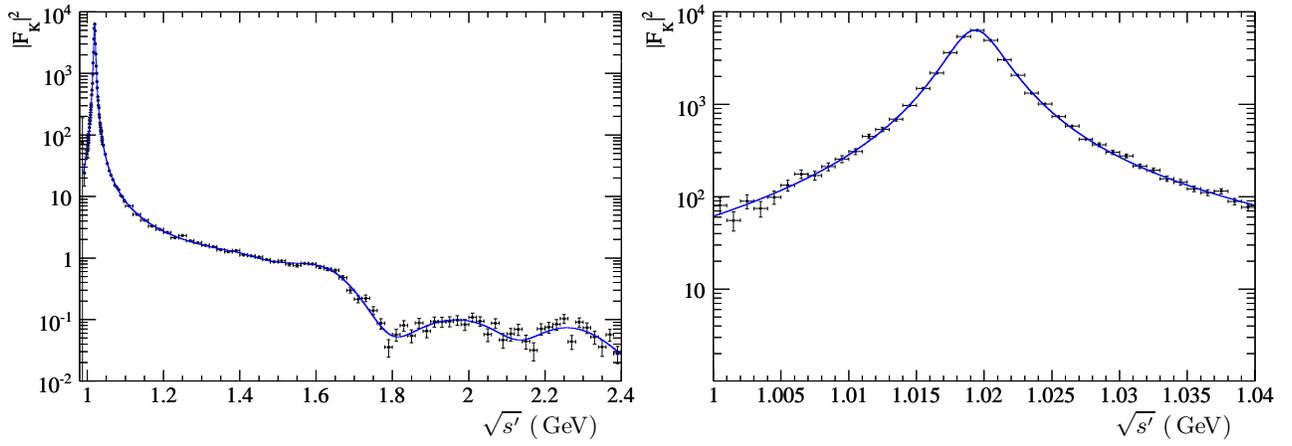

FIG. 26: Fit of the squared *BABAR* charged kaon form factor with a model based on a sum of resonances (see text), in the energy interval from threshold up to 2.4 GeV (left) and [1–1.04] GeV (right). Systematic and statistical uncertainties are shown for data points, i.e., the diagonal elements of the total covariance matrices.

Figure 28 shows the various contributions to the form factor in the $\phi$ mass region. The total contribution is dominated by the $\phi$ resonance with a small correction from the interference of the $\phi$ with the $\rho$ and $\omega$ amplitudes. Uncertainties are estimated from fit variations, by changing the number of parameters related to the higher-mass resonances or fixing the $\rho$ and $\omega$ amplitudes to their SU(3) values.

### D. $\phi$ resonance parameters

The $\phi$ mass obtained from the form factor fit is

$$m_\phi = (1019.51 \pm 0.02 \pm 0.05)\,\text{MeV}/c^2, \quad (15)$$

where the first uncertainty is statistical and the second is the total systematic error, which is dominated by the mass scale uncertainty. The small uncertainty on the $\phi$ mass due to the fit itself ($0.02\,\text{MeV}/c^2$) is included in the quoted uncertainty. The fitted $\phi$ width is

$$\Gamma_\phi = (4.29 \pm 0.04 \pm 0.07)\,\text{MeV}, \quad (16)$$

where the first uncertainty is statistical and the second accounts for the resolution uncertainty and includes the uncertainty due to the fit (0.04 MeV). These results are in good agreement with the world average values [19], which are $(1019.455 \pm 0.020)\,\text{MeV}/c^2$ and $(4.26 \pm 0.04)\,\text{MeV}$ for the $\phi$ mass and width, respectively.

The amplitude $a_\phi = 0.9938 \pm 0.0066$ is obtained from the fit. The product of the electronic width of the $\phi$ with its branching fraction into $K^+K^-$ is related to the fitted parameters through:

$$\Gamma_\phi^{ee} \times \mathcal{B}(\phi \to K^+K^-) = \frac{\alpha^2 \beta^3(s, m_K)}{324}\, \frac{m_\phi^2}{\Gamma_\phi}\, a_\phi^2\, C_{\text{FS}}, \quad (17)$$

where the Coulomb contribution is included in the $\phi \to K^+K^-$ decay width. The product defined in Eq. (17) is proportional to the integral of the cross section over the $\phi$ resonance peak, and is consequently independent of the experimental resolution. The form factor can indeed be directly expressed and fitted in terms of that product, with the result:

$$\Gamma_\phi^{ee} \times \mathcal{B}(\phi \to K^+K^-) = (0.6340 \pm 0.0070_{\text{exp}} \pm 0.0037_{\text{fit}} \pm 0.0013_{\text{cal}})\,\text{keV}, \quad (18)$$

where the first uncertainty is the total uncertainty (statistical plus systematic) on the cross section, the second is due to the fit, and the third is from the mass calibration. The result reported in Eq. (18) is the most precise from a single experiment. It is higher by 1.8 standard deviation of the combined errors compared to the most recent value extracted from CMD-2 [21] data: $(0.605 \pm 0.004 \pm 0.013)\,\text{keV}$.

It is not possible with the $K^+K^-$ *BABAR* data alone to separate $\Gamma_\phi^{ee}$ and $\mathcal{B}(\phi \to K^+K^-)$. The world aver-tween the rates from the two $K\bar{K}$ modes, which is well beyond the estimated isospin-breaking corrections [32].

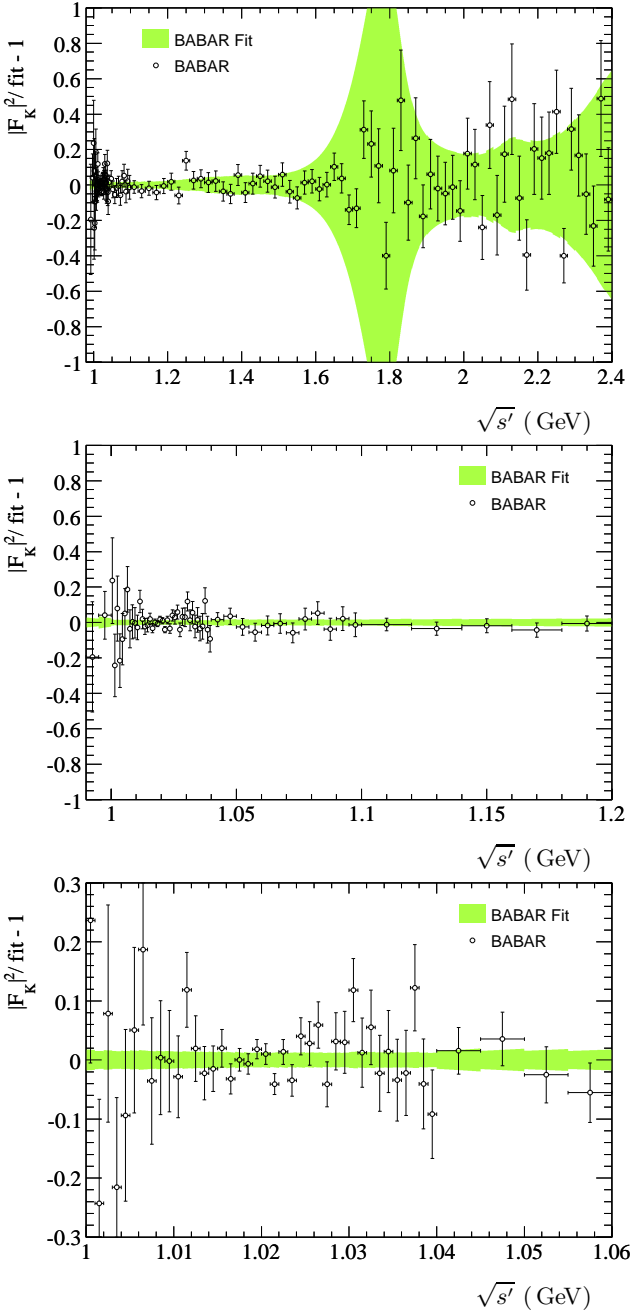

FIG. 27: Relative difference between the charged kaon squared form factor from *BABAR* data and the 19-parameter phenomenological fit in three mass regions. Systematic and statistical uncertainties are included for data (diagonal elements of the total covariance matrices). The width of the band shows the propagation of statistical errors in the fit and the quoted systematic uncertainties, added quadratically.

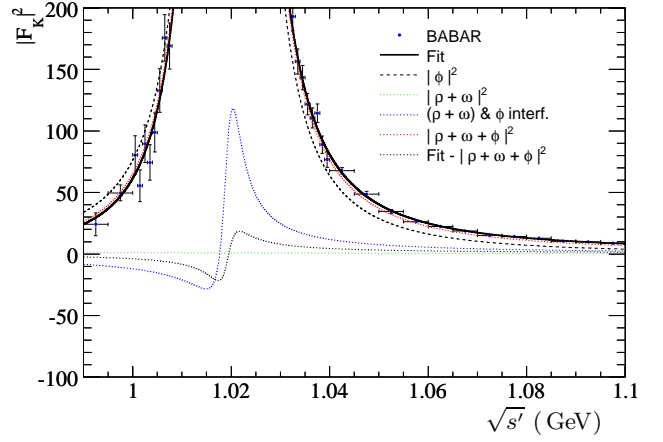

FIG. 28: (color online). Different contributions to the fit of the squared *BABAR* charged kaon form factor (black line) in the energy interval [0.99–1.1] GeV. The dominant contribution under the overwhelming $\phi$ resonance is from the interference between the $\rho + \omega$ and $\phi$ amplitudes (dotted blue line).

### E. Comparison to other $e^+e^-$ results

The measured form factor is compared to data published by previous experiments. Figure 29 shows the relative difference in the $\phi$ mass region between the *BABAR* fit and the CMD2 [21] and SND [22] data. While the uncertainty of the *BABAR* cross section at the $\phi$ is $7.2 \times 10^{-3}$ (Table I), systematic normalization uncertainties of 2.2% and 7.1% are reported by CMD2 and SND, respectively. In addition, the *BABAR* result, as well as the Novosibirsk measurements, are affected by systematic uncertainties on mass calibration, which are not included in Fig. 29. They amount to $0.08\,\mathrm{MeV}/c^2$ for both the CMD2 and SND experiments [21], fully correlated, and to $0.05\,\mathrm{MeV}/c^2$ for *BABAR*.

Differences observed in Fig. 29 are fitted assuming they result from differences in the $\phi$ mass calibration and normalization of the cross section through the quantities

$$\Delta m = m_\phi(BABAR) - m_\phi(\mathrm{CMD2, SND}),$$
$$\lambda = \frac{\mathrm{norm}(BABAR)}{\mathrm{norm}(\mathrm{CMD2, SND})} - 1 \quad (19)$$

The comparison between *BABAR* and CMD2 yields

$$\Delta m = (0.093 \pm 0.008_{(\mathrm{CMD2})} \pm 0.013_{(BABAR)})\,\mathrm{MeV}/c^2,$$
$$\lambda = 0.051 \pm 0.003_{(\mathrm{CMD2})} \pm 0.006_{(BABAR)}, \quad (20)$$

while the fit of the difference between *BABAR* and SND yields

$$\Delta m = (0.065 \pm 0.026_{(\mathrm{SND})} \pm 0.013_{(BABAR)})\,\mathrm{MeV}/c^2,$$
$$\lambda = 0.096 \pm 0.009_{(\mathrm{SND})} \pm 0.006_{(BABAR)}, \quad (21)$$

age values of these two quantities have been obtained from measurements of the four dominant $\phi$ decay modes $(K^+K^-, K_S K_L, \pi^+\pi^-\pi^0, \eta\gamma)$ by CMD2 and SND. When including the *BABAR* result on the $\Gamma_\phi^{ee} \times \mathcal{B}(\phi \to K^+K^-)$ product, one expects both $\Gamma_\phi^{ee}$ and $\mathcal{B}(\phi \to K^+K^-)$ to increase, thus reducing the long-standing discrepancy be-where only statistical uncertainties are included. The observed mass differences are compatible with the *BABAR* and CMD2 (SND) calibration uncertainties, but the normalization differences are not consistent by large factors with the quoted systematic uncertainties.

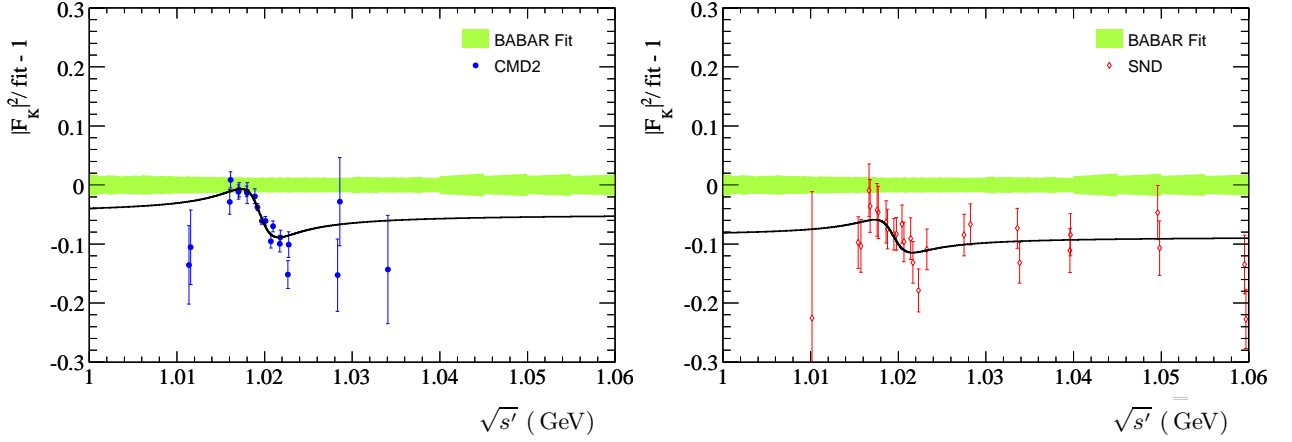

FIG. 29: Relative difference between the charged kaon squared form factor from CMD2 (left) and SND (right) data, and the *BABAR* phenomenological fit in the $\phi$ mass region. Only the statistical uncertainties are included for data (diagonal elements of the covariance matrix). The width of the band shows the propagation of statistical errors in the *BABAR* fit and the quoted systematic uncertainties, added quadratically. The solid line shows a fit of the relative difference, with $\phi$ masses different by $\Delta m$ (see text).

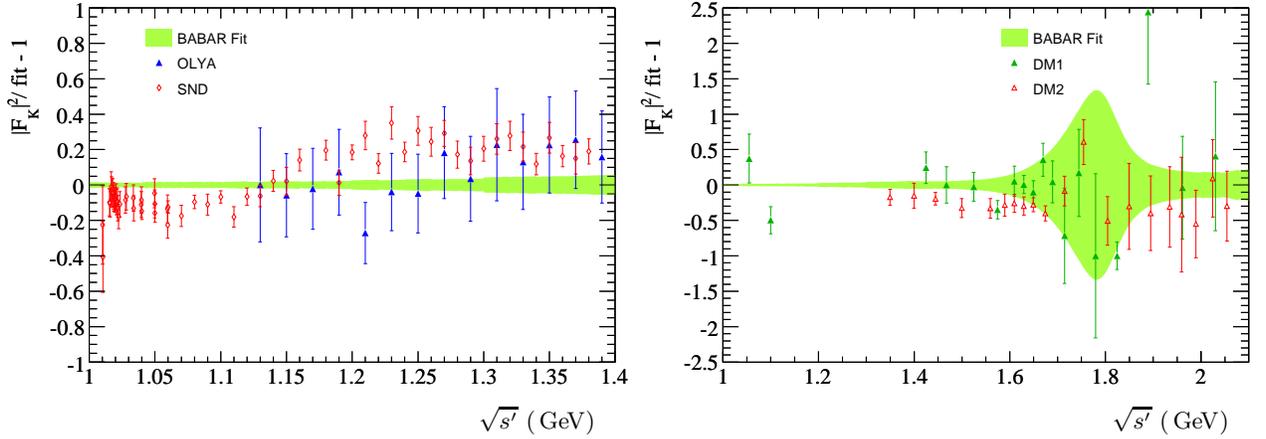

FIG. 30: Relative difference between the charged kaon squared form factor from SND and OLYA (left), and DM1 and DM2 (right), and the *BABAR* phenomenological fit in different mass regions. Systematic and statistical uncertainties are included for data (diagonal elements of the covariance matrix). The width of the band shows the propagation of statistical errors in the fit and the quoted systematic uncertainties, added quadratically.

The comparisons with the SND [23], OLYA [24], DM1 [25], and DM2 [26] measurements at higher masses are shown in Fig. 30. The systematic negative difference between *BABAR* and SND persists up to about 1.15 GeV, where a crossover occurs, while at higher masses, the SND values are consistently larger than the ones from *BABAR*. The *BABAR* data are in rather good agreement with data from OLYA and DM1, while a systematic difference is obtained when comparing to DM2.

## F. A fit to the *BABAR* form factor in the high mass region

The phenomenological fit to the *BABAR* form factor describes the data reasonably well up to 2.4 GeV. At higher masses, the form factor can be compared to the QCD prediction [33, 34] for its asymptotic behaviour:

$$F_K(s) = 16\pi\,\alpha_s\,(s)\,\frac{f_{K^+}^2}{s}. \qquad (22)$$

The result of the fit of the squared form factor between 2.5 and 5 GeV with the function $A\alpha_s^2(s)/s^n$ is shown in Fig. 31. $A$ and $n$ are left free in the fit, and the contributions of the narrow $J/\psi$ and $\psi(2S)$ resonances decaying to $K^+K^-$ are subtracted from the mass spectrum before performing the fit.

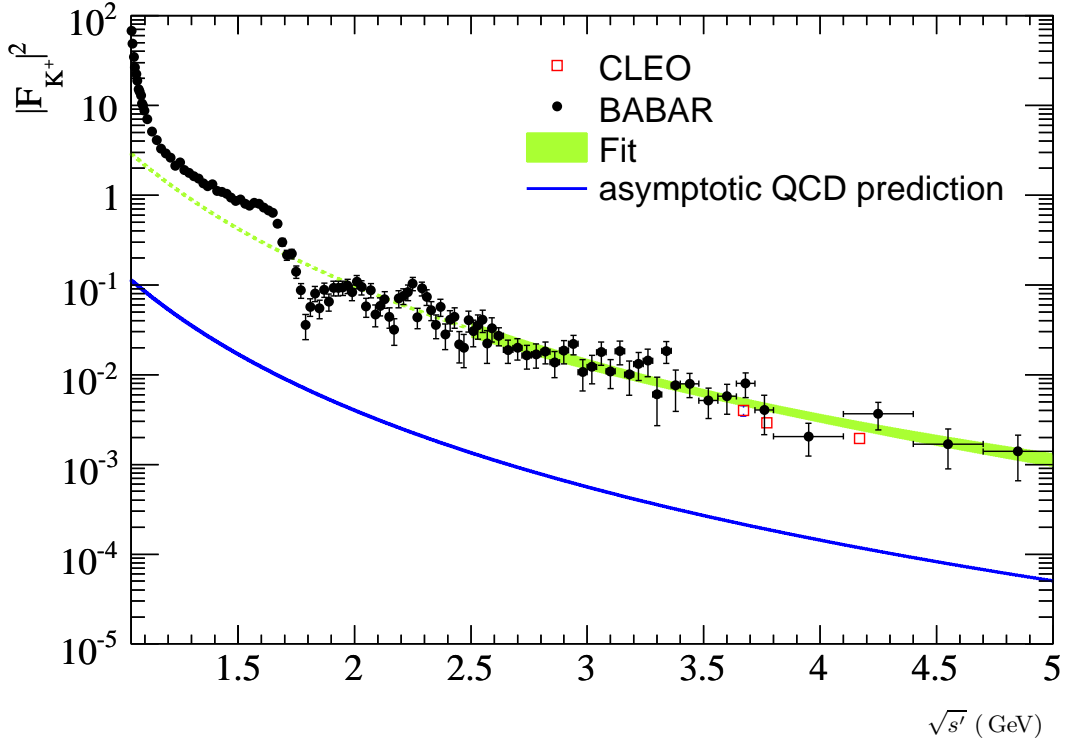

FIG. 31: (color online). Fit (green band) of the squared *BABAR* charged kaon form factor in the high mass region, using a function that has the shape of the QCD prediction (blue curve, see text). The extrapolation of the fit at low energy is indicated by the dotted green line. We also indicate measurements from CLEO data (red squares), close to the $\psi(2S)$ mass and above. Systematic and statistical uncertainties are shown for data points, i.e., the diagonal elements of the total covariance matrices.

The fit describes the data well ($\chi^2/\mathrm{DF} = 23.4/32$), with $n = 2.04 \pm 0.22$, which is in good agreement with the QCD prediction $n = 2$. When extrapolated to lower masses, the fit follows the average shape of the spectrum down to about 1.7 GeV. However, the fitted form factor is about a factor of 4 larger than the absolute perturbative QCD prediction of Eq. (22). This confirms the normalization disagreement observed with the CLEO measurements [35, 36] near the $\psi(2S)$ mass and above.

## G. The branching fractions of $J/\psi$ and $\psi(2S)$ to $K^+K^-$

Fig. 32 (left) shows the $K^+K^-$ mass spectrum in data in the $J/\psi$ region using a fine binning. The distribution is fitted with a Gaussian with free amplitude, width and mass, over a constant term for the continuum. The fit yields $51.4 \pm 8.2$ $J/\psi$ events, corresponding to an integrated cross section of $(0.00341 \pm 0.00055_{\mathrm{stat}} \pm 0.00006_{\mathrm{resolMCsyst}} \pm 0.00019_{\mathrm{syst}})$ nb·GeV, where the last uncertainty is from Table I, excluding the contribution from background (negligible for $J/\psi$). The Gaussian width $(6.1 \pm 1.7)$ MeV, where the quoted uncertainty is statistical only, is compatible with the MC resolution $(7.6 \pm 0.3)$ MeV in the same mass region. The fitted $J/\psi$ mass $(3097.2 \pm 1.4)$ MeV/$c^2$ is consistent with the world average [19] $(3096.916 \pm 0.011)$ MeV/$c^2$ within the statistical uncertainty of the fit.

As the background from misidentified $J/\psi \to \mu\mu$ peaks at a higher mass, no subtraction is performed and the integral over the $J/\psi$ resonance yields the product of the $J/\psi$ leptonic width by the $J/\psi \to K^+K^-$ branching fraction:

$$\Gamma^{ee}_{J/\psi} \times \mathcal{B}(J/\psi \to K^+K^-) = \frac{m^2_{J/\psi} \, N_{J/\psi \to K^+K^-}}{6\pi^2 \, dL^{\mathrm{eff}}_{\mathrm{ISR}}/d\sqrt{s'} \, \varepsilon_{J/\psi} \, C} \tag{23}$$

$$= (1.42 \pm 0.23_{\mathrm{stat}} \pm 0.08_{\mathrm{syst}}) \, \mathrm{eV}, \tag{24}$$

where $dL^{\mathrm{eff}}_{\mathrm{ISR}}/d\sqrt{s'}$ is the effective ISR luminosity dis- cussed in Sec. VII A, $\varepsilon_{J/\psi}$ is the full selection efficiency at the $J/\psi$ mass, and $C = 3.894 \times 10^{11}$ nb·MeV$^2$ is a conversion constant. The first, dominant, uncertainty in Eq. (24) is statistical, while the second one is systematic.

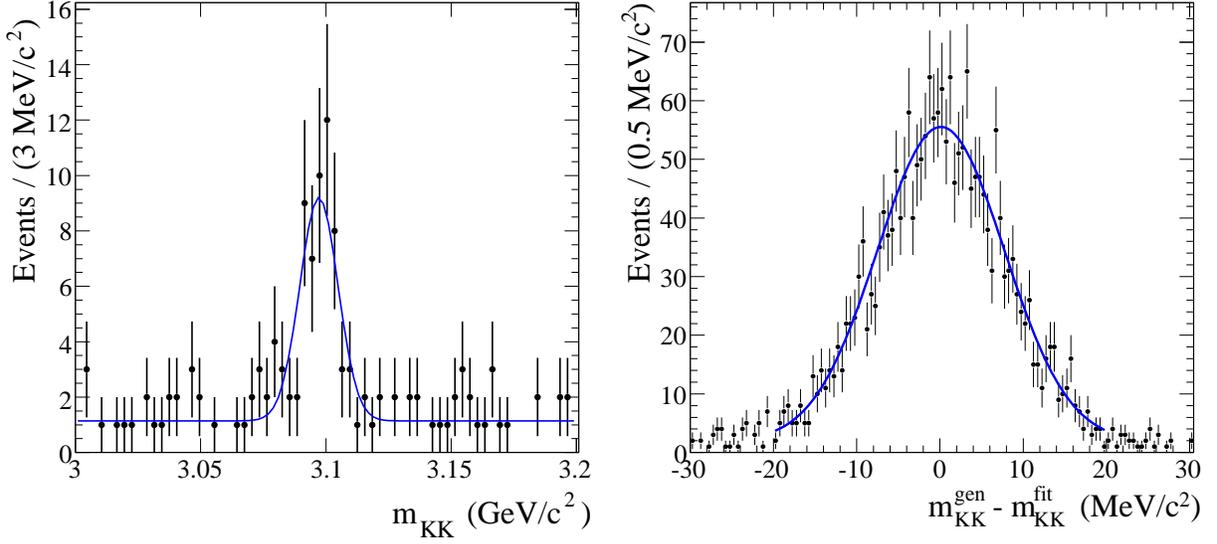

FIG. 32: Left: $K^+K^-$ mass spectrum in the data in the $J/\psi$ resonance region. Right: distribution of the difference between the generated and the fitted $K^+K^-$ mass in MC, for events with a generated mass between 3 and 3.2 GeV/$c^2$. The solid lines represent the results of fits by a Gaussian plus a constant term.

Using the precise world average value [19] for the leptonic width, $\Gamma^{ee}_{J/\psi} = (5.55 \pm 0.14)$ keV, one can deduce the $K^+K^-$ branching fraction:

$$\mathcal{B}(J/\psi \to K^+K^-) = (2.56 \pm 0.44_{\rm exp} \pm 0.07_{\Gamma^{ee}}) \times 10^{-4}, \quad (25)$$

in agreement with the world average value $(2.37 \pm 0.31) \times 10^{-4}$, dominated by the Mark-III result [37].

The same analysis is repeated for the weaker $\psi(2S)$ signal (Fig. 33). Using the MC resolution of $(9.2 \pm 1.1)$ MeV/$c^2$, the fit yields $10.8 \pm 4.2$ $\psi(2S)$ events, corresponding to an integrated cross section over the resonance of $(0.000596 \pm 0.000229_{\rm stat} \pm 0.000032_{\rm resolMCsyst} \pm 0.000034_{\rm syst})$ nb·GeV, where the last uncertainty is taken from Table I. The fitted $\psi(2S)$ mass $(3684.2\pm4.3)$ MeV/$c^2$ is consistent with the world average [19], $(3686.09 \pm 0.04)$ MeV/$c^2$, within the statistical uncertainty.

The integral over the resonance yields the product of the $\psi(2S)$ leptonic width times the $\psi(2S) \to K^+K^-$ branching fraction:

$$\Gamma^{ee}_{\psi(2S)} \times \mathcal{B}(\psi(2S) \to K^+K^-) = (0.35 \pm 0.14_{\rm stat} \pm 0.03_{\rm syst}) \, \text{eV}, \quad (26)$$

where the systematic error includes the uncertainty on the MC resolution width. Using the world average [19] for the leptonic width, $\Gamma^{ee}_{\psi(2S)} = (2.35 \pm 0.04)$ keV, one can deduce the $K^+K^-$ branching fraction:

$$\mathcal{B}(\psi(2S) \to K^+K^-) = (1.50 \pm 0.59_{\rm exp} \pm 0.03_{\Gamma^{ee}}) \times 10^{-4}, \quad (27)$$

in agreement with the world average value, $(0.63\pm0.07)\times 10^{-4}$.

## H. The $K^+K^-$ contribution to the anomalous magnetic moment of the muon

The bare $e^+e^- \to K^+K^-(\gamma)$ cross section obtained in this analysis can be used to compute the contribution of the $K^+K^-$ mode to the theoretical prediction of the anomalous magnetic moment of the muon.

The result of the dispersion integral is

$$a^{KK,{\rm LO}}_\mu = (22.93 \pm 0.18_{\rm stat} \pm 0.22_{\rm syst} \pm 0.03_{\rm VP}) \times 10^{-10}, \quad (28)$$

for the energy interval between the $K^+K^-$ production threshold and 1.8 GeV. The first uncertainty is statistical, the second is the experimental systematic, while the third is from the $\phi$ parameters used in the VP correction (Sec. VII A). The precision achieved is 1.2%, with systematic uncertainties contributing most to the total error. This is the most precise result for the $K^+K^-$ channel, and the only one covering the full energy range of interest. For comparison, the combination of all previous data [38] for the same range is $(21.63 \pm 0.27_{stat} \pm 0.68_{syst}) \times 10^{-10}$.

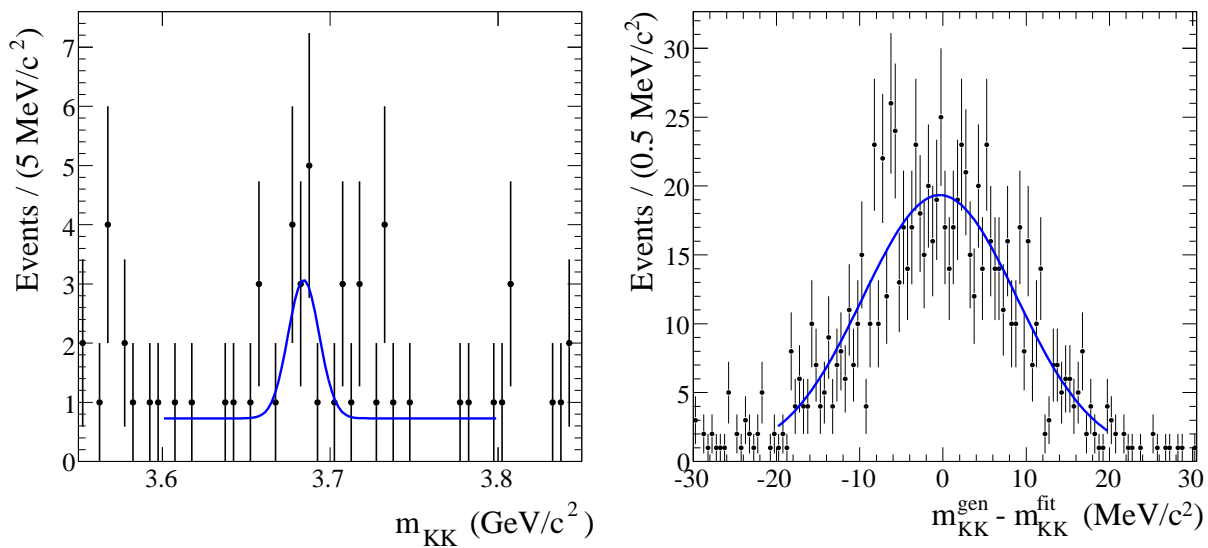

FIG. 33: Left: $K^+K^-$ mass spectrum in data in the $\psi(2S)$ resonance region. Right: distribution of the difference between the generated and the fitted $K^+K^-$ mass in MC, for events with a generated mass between 3.6 and 3.8 GeV/$c^2$. The solid lines represent the results of fits by a Gaussian plus a constant term.

While the choice of the upper integration limit is arbitrary, the value of 1.8 GeV is chosen as a convenient and practical transition [39, 40] between data and perturbative QCD in the dispersion integral. The $K^+K^-$ contribution in the range [1.8–3.0] GeV from the present measurement is only $(0.121 \pm 0.003_{stat} \pm 0.008_{syst}) \times 10^{-10}$. The quoted result Eq. (28) is dominated by the $\phi$ region, with a contribution of $(18.64 \pm 0.16_{stat} \pm 0.13_{syst} \pm 0.03_{VP}) \times 10^{-10}$ from threshold to 1.06 GeV.

## VIII. CONCLUSION

The cross section for the process $e^+e^- \to K^+K^-(\gamma)$ has been measured by the *BABAR* experiment, from the $K^+K^-$ production threshold to 5 GeV. The measurement uses the ISR method and the effective ISR luminosity determined with the $\mu^+\mu^-(\gamma)\gamma_{ISR}$ events in the same sample, as developed for the precision measurement of the $e^+e^- \to \pi^+\pi^-(\gamma)$ cross section [5].

The cross section is obtained for the first time continuously over the full energy range, with an overall systematic uncertainty of $7.2 \times 10^{-3}$ in the [1.01–1.03] GeV mass range. It spans more than six orders of magnitude and is dominated by the $\phi$ resonance close to threshold. Other structures visible at higher masses include the contributions from the narrow $J/\psi$ and $\psi(2S)$ resonances, which have been studied explicitly.

A fit of the charged kaon form factor has been performed using a sum of contributions from isoscalar and isovector vector mesons: besides the dominant $\phi$ resonance and small $\rho$ and $\omega$ contributions, several higher states are needed to reproduce the structures observed in the measured spectrum. Precise results for the mass and width of the $\phi$ resonance have been determined, and are found to agree with the world average values.

The results are in agreement with previous data at large energy and confirm the large normalization disagreement with the asymptotic QCD expectation already observed by the CLEO experiment. In the $\phi$ region, discrepancies with CMD-2 and SND results are observed in the normalization of the cross section. The differences exceed the uncertainties quoted by either experiment.

Finally, the *BABAR* results are used as input to the dispersion integral yielding the $K^+K^-(\gamma)$ vacuum polarization contribution at LO to the muon magnetic anomaly. This contribution amounts to $(22.93 \pm 0.18_{stat} \pm 0.22_{syst} \pm 0.03_{VP}) \times 10^{-10}$, dominated by the $\phi$ region.

We are grateful for the extraordinary contributions of our PEP-II colleagues in achieving the excellent luminosity and machine conditions that have made this work possible. The success of this project also relies critically on the expertise and dedication of the computing organizations that support *BABAR*. The collaborating institutions wish to thank SLAC for its support and the kind hospitality extended to them. This work is supported by the US Department of Energy and National Science Foundation, the Natural Sciences and Engineering Research Council (Canada), the Commissariat à l'Energie Atom-ique and Institut National de Physique Nucléaire et de Physique des Particules (France), the Bundesministerium für Bildung und Forschung and Deutsche Forschungsgemeinschaft (Germany), the Istituto Nazionale di Fisica Nucleare (Italy), the Foundation for Fundamental Research on Matter (The Netherlands), the Research Council of Norway, the Ministry of Education and Science of the Russian Federation, Ministerio de Economía y Competitividad (Spain), and the Science and Technology Facilities Council (United Kingdom). Individuals have received support from the Marie-Curie IEF program (European Union) and the A. P. Sloan Foundation (USA).

[1] V.N. Baier and V.S. Fadin, Phys. Lett. B **27**, 223 (1968).

[2] A.B. Arbuzov *et al.*, J. High Energy Phys. **9812**, 009 (1998).

[3] S. Binner, J.H. Kühn, and K. Melnikov, Phys. Lett. B **459**, 279 (1999).

[4] M. Benayoun *et al.*, Mod. Phys. Lett. A **14**, 2605 (1999).

[5] B. Aubert *et al.* (*BABAR* Collaboration), Phys. Rev. Lett. **103**, 231801 (2009); J.P. Lees *et al.* (*BABAR* Collaboration), Phys. Rev. D **86**, 032013 (2012).

[6] B. Aubert *et al.* (*BABAR* Collaboration), Nucl. Instr. Meth. A **479**, 1 (2002).

[7] H. Czyż and J.H. Kühn, Eur. Phys. J. C **18**, 497 (2001).

[8] M. Caffo, H. Czyż, and E. Remiddi, Nuovo Cim. **110A**, 515 (1997).

[9] E. Barberio, B. van Eijk, and Z. Was, Comput. Phys. Comm. **66**, 115 (1991).

[10] T. Sjöstrand, Comput. Phys. Comm. **82**, 74 (1994).

[11] S. Jadach and Z. Was, Comput. Phys. Comm. **85**, 453 (1995).

[12] S. Agostinelli *et al.*, Nucl. Instr. Meth. A **506**, 250 (2003).

[13] H. Czyż *et al.*, Eur. Phys. J. C **35**, 527 (2004); Eur. Phys. J. C **39**, 411 (2005).

[14] B. Aubert *et al.* (*BABAR* Collaboration), arXiv:1305.3560 [physics.ins-det] (in print in Nucl. Instr. Meth. A).

[15] B. Aubert *et al.* (*BABAR* Collaboration), Phys. Rev. D **77**, 092002 (2008).

[16] J.P. Lees *et al.* (*BABAR* Collaboration), Phys. Rev. D **86**, 012008 (2012).

[17] B. Aubert *et al.* (*BABAR* Collaboration), Phys. Rev. D **70**, 072004 (2004).

[18] B. Aubert *et al.* (*BABAR* Collaboration), Phys. Rev. D **73**, 012005 (2006).

[19] K. Nakamura *et al.* (Particle Data Group), J. Phys. **G37**, 075021 (2010).

[20] See EPAPS Document No.xxxx for files containing the cross section and form factor values and the covariance matrices. For more information on EPAPS, see http://www.aip.org/pubservs/epaps.html.

[21] R.R. Akhmetshin *et al.* (CMD-2 Collaboration), Phys. Lett. B **669**, 217 (2008).

[22] M.N. Achasov *et al.* (SND Collaboration), Phys. Rev. D **63**, 072002 (2001).

[23] M.N. Achasov *et al.* (SND Collaboration), Phys. Rev. D **76**, 072012 (2007).

[24] P.M. Ivanov *et al.* (OLYA Collaboration), Phys. Lett. B **107**, 297 (1981); P.M. Ivanov *et al.* (OLYA Collaboration), JETP Lett. **36**, 112 (1982).

[25] B. Delcourt *et al.* (DM1 Collaboration), Phys. Lett. B **99**, 257 (1981); F. Mané *et al.* (DM1 Collaboration), Phys. Lett. B **99**, 261 (1981).

[26] D. Bisello *et al.* (DM2 Collaboration), Z. Phys. C **39**, 13 (1988).

[27] Yu. M. Bystritskiy *et al.*, Phys. Rev. D **72**, 114019 (2005).

[28] H. Czyż *et al.*, Eur. Phys. J. C **39**, 411 (2005).

[29] A. Hoefer, J. Gluza, and F. Jegerlehner, Eur. Phys. J. C **24**, 51 (2002).

[30] C. Bruch, A. Khodjamirian, and J. H. Kuehn, Eur. Phys. J. C **39**, 41 (2005).

[31] G. D'Agostini, Nucl. Inst. Meth. A **346**, 306 (1994).

[32] A. Bramon *et al.*, Phys. Lett. B **486**, 406 (2000).

[33] V.L. Chernyak, A.R. Zhitnitsky, and V.G. Serbo, JETP Lett. **26**, 594 (1977).

[34] G.P. Lepage and S.J. Brodsky, Phys. Lett. B **87**, 359 (1979).

[35] T.K. Pedlar *et al.* (CLEO Collaboration), Phys. Rev. Lett. **95**, 261803 (2005).

[36] K. K. Seth *et al.*, Phys. Rev. Lett. **110**, 022002 (2013).

[37] R. M. Baltrusaitis *et al.*, Phys. Rev. D **32**, 566

[38] M. Davier, A. Hoecker, B. Malaescu, and Z. Zhang, Eur. Phys. J. C **71**, 1515 (2011).

[39] M. Davier and A. Hoecker, Phys. Lett. B **419**, 419 (1998).

[40] M. Davier, S. Eidelman, A. Hoecker, and Z. Zhang, Eur. Phys. J. C **27**, 497 (2003); C **31**, 503 (2003).